\documentclass[11pt,showpacs,preprint,preprintnumbers,nofootinbib,
superscriptaddress,amsmath,amssymb]{revtex4}

\pdfoutput=1
\usepackage{graphicx}
\usepackage{dcolumn}
\usepackage{bm}
\usepackage{amssymb}
\usepackage{amsmath}
\usepackage{epsfig}    
\usepackage{color}
\usepackage{slashed}

\newcommand{\GeV}{\textrm{GeV}}
\newcommand{\PI}{\textrm{PI}}
\newcommand{\RE}{\textrm{Re}}
\newcommand{\EM}{\textrm{em}}

\newcommand{\SM}{\textrm{SM}}
\newcommand{\HSM}{\textrm{HSM}}

\begin{document} 

\title{Radiative corrections to the Higgs boson couplings in the model with an additional real singlet scalar field}

\preprint{UT-HET 105}
\author{Shinya Kanemura}
\affiliation{Department of Physics, University of Toyama, \\3190 Gofuku, Toyama 930-8555, Japan}
\author{Mariko Kikuchi}
\affiliation{Department of Physics, University of Toyama, \\3190 Gofuku, Toyama 930-8555, Japan}
\author{Kei Yagyu}
\affiliation{School of Physics and Astronomy, University of Southampton, \\Southampton, SO17 1BJ, United Kingdom}

\begin{abstract}
We calculate renormalized Higgs boson couplings with gauge bosons and fermions at the one-loop level 
in the model with an additional isospin singlet real scalar field.
These coupling constants can deviate from the predictions in the standard model due to tree-level mixing effects and one-loop contributions of the extra  neutral scalar boson. 
We investigate how they can be significant under the theoretical constraints from perturbative unitarity and vacuum stability and also 
the condition of avoiding the wrong vacuum.
Furthermore, comparing with the predictions 
in the Type I two Higgs doublet model, 
we numerically demonstrate how the singlet extension model can be distinguished and identified by using precision measurements of the Higgs boson couplings 
at future collider experiments.
 \end{abstract}
\maketitle
\clearpage

\section{Introduction}

Although a Higgs boson was discovered by the LHC experiments in 2012~\cite{LHC_ATLAS, LHC_CMS}, the structure of the Higgs sector and the physics behind the Higgs sector remain unknown.
Deep understanding for the Higgs sector is a key to explore new physics beyond the standard model (SM).

The minimal Higgs sector of the SM satisfies the current LHC data~\cite{LHC_ATLAS2, LHC_CMS2}, 
while most of non-minimal Higgs sectors do so as well.
As there is no theoretical reason to choose the minimal form of the Higgs sector like in the SM, 
there are many possibilities for extended Higgs sectors which contain additional scalar isospin multiplets. 
In principle, there are infinite kinds of extended Higgs sectors.
However, particular importance exists in the second simplest Higgs sectors, where only one isospin multiplet is added to the SM Higgs sector,
such as a model with an additional singlet, doublet or triplet scalar field.  
There are many new physics models which predict one of these extended Higgs sectors such as 
the $B-L$ extended SM with the $B-L$ symmetry breaking~\cite{B-L} which contains an additional singlet scalar field, 
the minimal supersymmetric SM~\cite{MSSM, Higgs_hunters} whose Higgs sector has the two Higgs doublets,  
and the model for the Type II seesaw mechanism~\cite{HTM_models} which can generate Majorana neutrino masses by introducing a complex triplet Higgs field, and so on.
These second simplest Higgs sectors can also be regarded as low-energy effective theories of more complicated Higgs sectors.

How can we test extended Higgs sectors by experiments?
Obviously direct discovery of the second Higgs boson is manifest evidence for extended Higgs sectors. 
By detailed measurements of such a particle, we can determine the structure of the Higgs sector.
On the other hand, we can also test extended Higgs sectors by precisely measuring low energy observables such as those in flavor physics~\cite{Flavour}, electroweak precision observables~\cite{LEP}, etc. 
As additional observables we can consider a set of coupling constants of the discovered Higgs boson. 
In general, a pattern of deviations in these observables strongly depends on the effects of extra Higgs bosons and other new physics particles, 
so that we may be able to fingerprint extended Higgs sectors and new physics models if we can detect a special pattern of the deviations at future experiments~\cite{2HDM_yukawa_reno,exotic, HTM_reno1,2HDM_full,FPs1,FPs2,white_paper}.

After the Higgs boson discovery, coupling constants of the discovered Higgs boson with SM particles became new observables 
to be measured as precisely as possible at current and future colliders.
Currently the measured accuracies for the Higgs boson couplings are typically of the order of 10 \%~\cite{LHC_ATLAS2, LHC_CMS2}. 
They will be improved drastically to the order of 1 \% or even better at future lepton colliders, such as the International Linear Collider (ILC)~\cite{ref:ILC_TDR,white_paper}, 
the Compact LInear Collider (CLIC)~\cite{CLIC} 
and Future $e^+e^-$ Circular Collider (FCCee). 
Therefore, these future electron-positron colliders are idealistic tools for fingerprinting Higgs sector and new physics models via precise measurements of the Higgs boson couplings.
In order to compare theory predictions with such precision measurements, 
calculations with higher order corrections are clearly necessary.

One-loop corrected Higgs boson couplings have been calculated in two Higgs doublet models (THDMs) with various Yukawa interactions~\cite{2HDM_full,KOSY,2HDM_yukawa_reno}, in the inert doublet model (IDM)~\cite{Arhrib_IDM, KKS} and in the Higgs triplet model (HTM)~\cite{HTM_reno1,HTM_reno2, HTM_reno3}.
In addition, decay rate of loop induce processes $hgg$, $h\gamma\gamma$ and $h\gamma Z$ have been investigated in THDMs~\cite{2HDM_hgammagamma, 2HDM_hgammagamma_hZg, 2HDM_hgammagamma_hbb,2HDM_3HDM_hgammagamma}, the IDM~\cite{IDM_loop} and the HTM~\cite{triplet_hgammagamma,triplet_hZg,triplet_hgammagamma_hZg,extended_hgammagamma_hZg}.

In this paper, we calculate one-loop corrections to the Higgs boson couplings with gauge bosons and with fermions
in the model with a real isospin singlet scalar field (HSM)~\cite{HSM_pheno,EWBG,Fuyuto,di_Higgs_HSM,chen_dawson_lewis,di_Higgs_QCD_HSM}. 
The renormalized couplings can deviate from the SM predictions due to the mixing effect and the one-loop contributions of the extra neutral scalar boson. 
The one-loop contributions are calculated in the on-shell scheme. 
We numerically investigate how they can be significant under the theoretical constraints from perturbative unitarity and vacuum stability and also 
the conditions of avoiding the wrong vacuum. 
Furthermore, we compare the results with the predictions at the one-loop level
in the THDM with Type I Yukawa interactions~\cite{2HDM_full,2HDM_yukawa_reno}. 
We study how the HSM can be distinguished from these models and identified by using precision measurements of the Higgs boson couplings 
at future collider experiments.

This paper is organized as follows.
In Sec.~II, we define the HSM, and briefly review the tree level properties to fix notation.
In Sec.~III, we present our calculational scheme for one-loop corrections to the Higgs boson couplings in the HSM. 
Sec.~IV is devoted to showing the numerical results of the renormalized scaling factors of the Higgs boson couplings. 
In Sec.~V, we show deviations in the $hZZ$, $hb\bar{b}$ and $h\gamma\gamma$ couplings in the HSM together with those 
in the THDM with the Type I Yukawa interaction, 
and see how we can discriminate these models by using future precision measurements of these couplings. 
Conclusions are given in Sec.~VI.
Various formulae are collected in Appendix.

\section{Model}\label{sec2}

The scalar sector of the HSM is composed of a complex isospin doublet field $\Phi$ with hypercharge $Y=1/2$ and a real singlet field $S$ with $Y=0$. 
The most general Higgs potential is given by
 \begin{align}
 V(\Phi,S) &= m_\Phi^2 |\Phi|^2 + \lambda |\Phi|^4 + \mu_{\Phi S}^{}|\Phi|^2 S
 + \lambda_{\Phi S}^{}|\Phi|^2 S^2 
 + t_S^{} S + m_S^2 S^2 + \mu_S S^3 + \lambda_S S^4,
 \label{pote}
 \end{align}
where all parameters are real.
The Higgs fields $\Phi$ and $S$ can be parametrized,
 \begin{align}
 \Phi =\left(\begin{array}{c}
             G^+ \\
             \frac{1}{\sqrt{2}}(\phi + v +i G^0) \\
             \end{array}\right), \,\,
 S = s + v_S^{},
 \end{align}
where $v$ and $v_S^{}$ are vacuum expectation values (VEVs) of $\Phi$ and $S$, respectively.
The fields $G^+$ and $G^0$ are Nambu-Goldstone bosons to be absorbed in longitudinally polarized weak gauge bosons.
Notice that $v$ is determined by the Fermi constant $G_F^{}$ by $v=1/(\sqrt{2}G_F)^{1/2}$ ($\simeq$ 246 GeV) while $v_S^{}$ does not affect electroweak symmetry breaking.
As it has been pointed it out in Refs.~\cite{wrong_vacuum, chen_dawson_lewis}, 
the potential in Eq.~(\ref{pote}) is invariant under the transformation of $v_S^{}\to v_S'$ by redefining 
all the potential parameters associated with $S$.

At the tree level, tadpoles are given by 
 \begin{align}
& T_\phi = v \{m_\Phi^2 + \lambda v^2 + v_S^{} (\lambda_{\Phi S} v_S^{} + \mu_{\Phi S}^{})
             \},\\
& T_s = t_S^{} + 2 m_S^2 v_S^{} + 4 \lambda_S v_S^3 + \lambda_{\Phi S} v^2 v_S^{} + 
 3 \mu_S^{} v_S^2 + \frac{\mu_{\Phi S}v^2}{2}.
 \label{tad pole}
 \end{align}
By imposing the stationary condition $T_\phi = 0$ and $T_s = 0$, $m_\Phi^2$ and $t_S^{}$ are related to the other parameters as
 \begin{align}
 & m_\Phi^2 = -\lambda v^2 - \lambda_{\Phi S} v_S^2 - \mu_{\Phi S} v_S, 
  \label{eq:mphi}\\
 & t_S^{} = - 2m_S^2 v_S^{} -4 \lambda_S^{} v_S^3 
         - \lambda_{\Phi S}^{} v_S^{} v^2
         - 3 v_S^2 \mu_S^{} - \frac{1}{2}v^2 \mu_{\Phi S}^{}.\label{eq:ts} 
 \end{align}

 After the electroweak symmetry breaking, mass terms of the scalar fields can be expressed as
 \begin{align}
 \mathcal{L}_{\textrm{mass}}^{} =
 -\frac{1}{2} (s ,  \phi )
  \left( \begin{array}{cc}
         M_{11}^{2} & M_{12}^{2} \\
         M_{12}^{2} & M_{22}^{2} \\
         \end{array} \right)
  \left( \begin{array}{c}
          s \\
          \phi \\
          \end{array} \right),
 \end{align}
 where
 \begin{align}
 & M_{11}^2 =  \mathcal{M}^2 + \lambda_{\Phi S}^{}v^2,\\
 & M_{12}^2 = (2\lambda_{\Phi S} v_S^{} + \mu_{\Phi S}^{})v, \\
 & M_{22}^2 =  2\lambda v^2,  
 \end{align} 
 with 
 \begin{align}
 \mathcal{M}^2 = 2m_S^2 + 12 \lambda_{S}^{}v_S^2 + 6 \mu_S^{}v_S^{}.
 \end{align}
We diagonalize the mass matrix by introducing the mixing angle $\alpha$, 
and express the scalar fields by mass eigenstates $H$ and $h$, 
 \begin{align}
 \mathcal{L}_{\textrm{mass}} =  -\frac{1}{2} (H,h)
  \left( \begin{array}{cc}
         m_H^{2} & 0 \\
         0 & m_h^{2} \\
         \end{array} \right)
  \left( \begin{array}{c}
          H \\
          h \\
          \end{array} \right),
 \end{align}
where the mass eigenstates $H$ and $h$ are related to the original fields  $s$ and $\phi$ by
 \begin{align}
 \left(\begin{array}{c}
       s \\
       \phi \\
        \end{array}\right) = \left(\begin{array}{cc} 
                 \cos\alpha & -\sin\alpha \\
                 \sin\alpha & \cos\alpha \\
                 \end{array} \right)
                                 \left(\begin{array}{c} 
                                            H \\
                                            h \\ 
                                           \end{array}\right).
 \end{align}
The masses of  $H$ and $h$ are given by
 \begin{align}
& m_H^2 = \cos^2\alpha M_{11}^2 + 
          \sin^2\alpha M_{22}^2 + \sin(2\alpha) M_{12}^2, \\
& m_h^2 = \sin^2\alpha M_{11}^2 + 
          \cos^2\alpha M_{22}^2 - \sin(2\alpha) M_{12}^2, 
 \end{align}
where $h$ identified to be the discovered Higgs boson with $m_h^{}\simeq 125$ GeV. 
The mixing angle $\alpha$ can be written in terms of the parameters in the potential as 
 \begin{align}
 \tan(2\alpha) = 
 \frac{2v(2\lambda_{\Phi S}^{} v_S^{} + \mu_{\Phi S}^{})}{\mathcal{M}^2 - v^2(2\lambda - \lambda_{\Phi S})}.
 \end{align}
We note that the SM limit is realized by taking $\mathcal{M}^2$ to be infinity.
In the following discussion, we use $s_\alpha$ and $c_\alpha$ to express $\sin\alpha$ and $\cos\alpha$, respectively.

By using physical parameters $m_h^2, m_H^2$ and $\alpha$,  
the three  parameters in the potential, $\lambda$, $m_S^2$, and $\mu_{\Phi S}^{}$, can be expressed as
 \begin{align}
 & \lambda = \frac{1}{2v^2}(c_\alpha^2 m_h^2 + s_\alpha^2 m_H^2), 
   \label{eq:lambda}\\
 & m_S^2 = \frac{c_\alpha^2 m_H^2}{2} + \frac{s_\alpha^2 m_h^2}{2}
    - 6\lambda_S^{} v_S^2 -\frac{1}{2}\lambda_{\Phi S}^{}v^2 -3v_S^{}\mu_S^{}, \label{eq:ms}\\
 & \mu_{\Phi S}^{} = \frac{s_{2\alpha}^{}}{2v}(m_H^2-m_h^2)-2\lambda_{\Phi S}^{}v_S^{}.
    \label{eq:mups}
 \end{align}
There are eight real parameters in the Higgs potential $m_\Phi^2$, $\lambda$, $\mu_{\Phi S}^{}$, $\lambda_{\Phi S}^{}$, $t_S^{}$, $m_S^2$, $\mu_S^{}$ and $\lambda_S^{}$, 
which are replaced by 
$v$, $m_h^2$, $m_H^2$, $\alpha$, $v_S$, $\lambda_{\Phi S}$, $\lambda_S^{}$ and $\mu_S^{}$.
  
The kinetic terms for the scalar fields are given by 
 \begin{align}
 \mathcal{L}_\textrm{kine} &= |D^\mu \Phi|^2 + \frac{1}{2}(\partial^\mu S)^2,
 \end{align} 
where $D^\mu = \partial^\mu -i \frac{g}{2} \tau_a^{} W_a^\mu - i \frac{g'}{2} B^\mu .$
We obtain interaction terms between weak gauge fields and scalar fields as
 \begin{align}
 \mathcal{L}_\textrm{kine} = 
    (s_\alpha H + c_\alpha h)\frac{2m_W^2}{v}g^{\mu\nu}W^+_\mu W^-_\nu
  + (s_\alpha H + c_\alpha h)\frac{m_Z^2}{v}g^{\mu\nu}Z_\mu Z_\nu
  + \cdots,
 \label{lag_gauge}
 \end{align}
where $m_W^{}$ and $m_Z^{}$ are the masses of $W$ and $Z$ bosons, respectively.
Although the Yukawa interaction is the same form as that in the SM,
Yukawa couplings are modified from the SM predictions by the field mixing,
 \begin{align}
 \mathcal{L}_{Y} = -\frac{m_f}{v} ( s_\alpha \bar{f}f H
                                 + c_\alpha \bar{f}f h), 
 \label{lag_yukawa}
 \end{align} 
where $m_f^{}$ represents the mass of a fermion $f$.

We define the scaling factors as ratios of the Higgs boson couplings in the HSM from those in the SM, 
 \begin{align}
 \kappa_V \equiv \frac{g_{hVV}^{\HSM}}{g_{hVV}^{\SM}}, \,\,\, 
  \textrm{for} \,\, V = W,Z, \,\,\,
 \kappa_f \equiv \frac{y_{hff}^{\HSM}}{y_{hff}^{\SM}}, \,\,\, 
 \kappa_h \equiv \frac{\lambda_{hhh}^{\HSM}}{\lambda_{hhh}^{\SM}},  
 \end{align} 
where $g_{hVV}^{\HSM(\SM)}, y_{hff}^{\HSM(\SM)}$ and $\lambda_{hhh}^{\HSM(\SM)}$ are coefficients of $hVV, hf\bar{f}$ and $hhh$ vertices in the HSM (SM), respectively.
Tree level values of $\kappa_V$, $\kappa_f$ and $\kappa_h$ are respectively derived from Eqs.~(\ref{lag_gauge}), (\ref{lag_yukawa}) and (\ref{pote}) as
 \begin{align}
 &\kappa_V = \kappa_f = c_\alpha^{}, \\
 & \kappa_h = c_\alpha^3 + \frac{2v}{m_h^2}s_\alpha^2(\lambda_{\Phi S} v c_\alpha - \mu_S s_\alpha
                             - 4s_\alpha \lambda_S^{} v_S^{}).
 \end{align}

We take into account several theoretical constraints in the HSM; i.e., perturbative unitarity~\cite{Uni}, vacuum stability~\cite{Fuyuto} and the condition to avoid a wrong vacuum~\cite{chen_dawson_lewis,wrong_vacuum}. 
We give the explanation of these theoretical constraints in Appendix \ref{sec:constraints}.

In addition to these theoretical constraints, the parameter space in the HSM is constrained by using experimental data. 
In Refs.~\cite{HSM_W, HSM_LHC-I}, the one-loop corrections to $m_W^{}$ has been calculated in the HSM with a discrete $Z_2$ symmetry. 
The limits on $s_\alpha$ and $m_H$ have been derived by comparing the prediction of $m_W$ and its measured value at the LEP experiment, namely, 
$|s_\alpha^{}| \gtrsim 0.3$ ($0.2$) with $m_H^{} = 300$ ($800$) GeV is excluded at the 2$\sigma$ level. 
Although the electroweak $S$, $T$ and $U$ parameters have also been calculated in Ref.~\cite{HSM_LHC-I}, constraints from those parameters 
are weaker than those from $m_W$.  
We show the formulae of one-loop corrected $m_W$ and also the $S$, $T$ and $U$ parameters in Appendix~\ref{sec:STU}.

 Null results from the Higgs boson searches at LEP and the LHC Run-I can constrain the signal rate of the second Higgs boson 
 which is defined as $S[H] \equiv \sigma[H] \times BR[H\rightarrow XY]$\footnote{Although we can obtain constraints on the signal rate of the additional Higgs boson by using the data at Tevetron, these constraints are entirely superseded by the one of the LHC Run-I~\cite{HSM_LHC-I}.}
 where $\sigma_H^{}$ and $BR[H \rightarrow XY]$ are the production cross section of $H$ and the branching fraction of the $H\rightarrow XY$ decay process in the HSM, respectively.
 If we assume $BR[H \rightarrow hh] =0$, $S[H]$ is given by $s_\alpha^2$ times the signal rate of the SM Higgs boson. 
In that case, constraints from LEP and LHC simply depend on $m_H^{}$ and $\alpha$. 
In Ref.~\cite{HSM_LHC-I}, the excluded parameter region on $m_H^{}$ and $\alpha$ has been presented using the LEP and the LHC results under the assumption of $BR[H \rightarrow hh] =0$.  
In the region with $m_H < 80$ GeV, most of the parameter regions of $\alpha$ have been excluded with 95 $\%$ CL.
In the region between 130 GeV and 500 GeV, $|s_\alpha^{}| \gtrsim 0.4$ is excluded with 95 $\%$ CL.
There are no constraints on $|s_\alpha^{}|$ for $m_H \gtrsim 800$ GeV.

\section{Renormalization}

In this section, we define the renormalization scheme of the HSM in order to calculate the one-loop corrected Higgs boson couplings. 
We describe how to determine each counter term in the gauge sector, the Yukawa sector and the Higgs sector.
We employ the same renormalization procedure as those given in Refs.~\cite{2HDM_yukawa_reno, 2HDM_full} for the gauge sector and the Yukawa sector, 
because the parameters in these sectors are exactly the same as those in the SM.

 \subsection{Renormalization in the gauge sector}\label{sec:reno gauge}

The gauge sector is described by three independent parameters as in the SM. 
When we choose $m_W^{}$, $m_Z^{}$ and $\alpha_\EM^{}$ 
as the input parameters, all the other parameters such as 
$v$ and weak mixing angle $\sin\theta_W^{}$ ($s_W^{}$) are given in terms of these three input parameters as
 \begin{align}
 & v^2= \frac{m_W^2}{\pi \alpha_\EM}\left(1-\frac{m_W^2}{m_Z^2}\right), \,\,\,\,\,  s_W^2 = 1-\frac{m_W^2}{m_Z^2}.
 \label{gauge rela1}
 \end{align} 
These parameters and weak gauge fields; namely $W^{\pm}_\mu, Z_\mu$ and $A_\mu$, are shifted as follows
 \begin{align}
 & m_W^2 \rightarrow m_W^2 + \delta m_W^2, \\
 & m_Z^2 \rightarrow m_Z^2 + \delta m_Z^2, \\
 & \alpha_\EM \rightarrow \alpha_\EM + \delta \alpha_\EM, \\
 & v \rightarrow v + \delta v, \\
 & s_W^2 \rightarrow s_W^2 + \delta s_W^2, \\ 
& W_\mu^\pm \rightarrow (1+\frac{1}{2}\delta Z_W) W_\mu^\pm, \\
 &\left(\begin{array}{c}
       A_\mu \\ 
       Z_\mu \\ 
       \end{array}\right) \rightarrow
 \left(\begin{array}{cc}
       1+\frac{1}{2}\delta Z_\gamma & 
       \frac{1}{2}\delta Z_{\gamma Z} +\frac{1}{2s_W c_W}\delta s_W^2 \\
       \frac{1}{2}\delta Z_{\gamma Z} -\frac{1}{2s_W c_W}\delta s_W^2 &
       1+\frac{1}{2}\delta Z_Z \\
       \end{array}\right)
 \left(\begin{array}{c}
       A_\mu \\
       Z_\mu \\
       \end{array}\right).
 \end{align}

Renormalized two point functions of gauge fields, $W^+W^-, ZZ, \gamma\gamma$ and the $\gamma Z$ mixing, are given by using the above counter terms and the 1PI diagrams denoted by $\Pi_{XY}^{1\PI}$ as
 \begin{align}
 \hat{\Pi}_{WW}[p^2]& =\Pi_{WW}^{1\PI}[p^2] +(p^2-m_W^2)\delta Z_W -\delta m_W^2,\\
 \hat{\Pi}_{ZZ}[p^2]& =\Pi_{ZZ}^{1\PI}[p^2] +(p^2-m_Z^2)\delta Z_Z -\delta m_Z^2,\\
 \hat{\Pi}_{\gamma\gamma}[p^2]& =\Pi_{\gamma\gamma}^{1\PI}[p^2] +p^2 \delta Z_\gamma,\\
 \hat{\Pi}_{\gamma Z}[p^2]& =\Pi_{\gamma Z}^{1\PI}[p^2] 
   -\frac{1}{2}(2p^2-m_Z^2)\delta Z_{\gamma Z} -\frac{m_Z^2}{2s_W c_W}\delta s_W^2,
 \end{align}
where
 \begin{align}
  \delta Z_W &= \delta Z_\gamma +\frac{c_W}{s_W}\delta Z_{\gamma Z}, \,\,\,\,\,
 \delta Z_{\gamma Z}^{} = 
                      \frac{s_W c_W}{c_W^2 - s_W^2}
                      (\delta Z_Z^{} - \delta Z_\gamma^{}).
 \label{gauge rela2}
 \end{align}
The explicit expressions of 1PI diagrams for gauge boson two point functions are given in Appendix.~\ref{sec:1PI 2point}.

Imposing following five renormalization conditions as~\cite{EW_reno}
 \begin{align}
 &\textrm{Re}\hat{\Pi}_{WW}[m_W^2]=0, \; 
 \textrm{Re}\hat{\Pi}_{ZZ}[m_Z^2]=0, \; 
 \hat{\Gamma}_{ee\gamma}^\mu[q^2=0,\slashed{p}_1=\slashed{p}_2=m_e^{}]= ie\gamma^\mu, \notag\\
 & \frac{d}{dp^2}\hat{\Pi}_{\gamma\gamma}[p^2]\Big|_{p^2=0} =0, \;
 \hat{\Pi}_{\gamma Z}[0] = 0,
 \end{align} 
five independent counter terms $\delta m_W^2, \delta m_Z^2, \delta \alpha_\EM, \delta Z_\gamma$ and $\delta Z_{\gamma Z}$ are determined as,
 \begin{align}
 \delta m_W^2 &= \textrm{Re}\Pi_{WW}^{1\PI}[m_W^2], \\
 \delta m_Z^2 &= \textrm{Re}\Pi_{ZZ}^{1\PI}[m_Z^2], \\ 
 \frac{\delta e}{e} &= \frac{1}{2} \frac{d}{dp^2}\Pi_{\gamma\gamma}^{1\PI}[p^2]\Big|_{p^2=0} - \frac{s_W}{c_W} \frac{\Pi_{\gamma Z}^{1\PI}[0]}{m_Z^2}, \\
 \delta Z_\gamma &=  -\frac{d}{dp^2}\Pi_{\gamma\gamma}^{1\PI}[p^2]\Big|_{p^2=0},\\
 \delta Z_{\gamma Z} &= -\frac{2}{m_Z^2}\Pi_{\gamma Z}^{1\PI}[0] + \frac{1}{s_W c_W}\delta s_W^2.
 \end{align}
Because of the relations Eqs.~(\ref{gauge rela1}) and (\ref{gauge rela2}),
other counter terms can be expressed by using above counter terms,
 \begin{align}
 \frac{\delta s_W^2}{s_W^2}& = 
    \frac{c_W^2}{s_W^2}\left(
    \frac{\delta m_Z^2}{m_Z^2} 
   -\frac{\delta m_W^2}{m_W^2} \right) \notag\\
   & =\frac{c_W^2}{s_W^2}\left(
    \frac{\RE\Pi_{ZZ}^{1\PI}[m_Z^2]}{m_Z^2} 
   -\frac{\RE\Pi_{WW}^{1\PI}[m_W^2]}{m_W^2} \right), \\
 \frac{\delta v}{v} &=\frac{1}{2}\left(
    \frac{\delta m_W^2}{m_W^2} -\frac{\delta \alpha_\EM}{\alpha_\EM}
    +\frac{\delta s_W^2 }{s_W^2}\right) \notag\\
  &= \frac{1}{2}\left(
   \frac{s_W^2-c_W^2}{s_W^2}\frac{\RE\Pi_{WW}^{1\PI}[m_W^2]}{m_W^2}
   + \frac{c_W^2}{s_W^2}\frac{\RE\Pi_{ZZ}^{1\PI}[m_Z^2]}{m_Z^2}
   - \frac{d}{dp^2}\Pi^{1\PI}_{\gamma\gamma}[p^2]\Big|_{p^2=0} -\frac{2s_W}{c_W}\frac{\Pi_{\gamma Z}^{1\PI}[0]}{m_Z^2}\right),
 \end{align}
 \begin{align}
 \delta Z_Z &= \frac{c_W^2-s_W^2}{s_W^2} 
    \left(\frac{\RE \Pi_{ZZ}^{1\PI}[m_Z^2]}{m_Z^2}
         -\frac{\RE \Pi_{WW}^{1\PI}[m_W^2]}{m_W^2} \right)
  - \frac{d}{dp^2}\Pi^{1\PI}_{\gamma\gamma}[p^2]\Big|_{p^2=0} +\frac{2}{m_Z^2}\frac{c_W^2-s_W^2}{s_W c_W}\Pi_{\gamma Z}^{1\PI}[0],
  \\
 \delta Z_W 
            &= \frac{c_W^2}{s_W^2}\left(
    \frac{\RE\Pi_{ZZ}^{1\PI}[m_Z^2]}{m_Z^2} 
   -\frac{\RE\Pi_{WW}^{1\PI}[m_W^2]}{m_W^2} \right)
   - \frac{d}{dp^2}\Pi^{1\PI}_{\gamma\gamma}[p^2]\Big|_{p^2=0} +\frac{2c_W}{s_W}\frac{\RE\Pi_{\gamma Z}[0]}{m_Z^2}.
 \end{align}

Because we have obtained explicit forms of counter terms in the gauge sector, we can calculate the one-loop level predictions for electroweak observables such as electroweak precision parameter $\Delta r$ and renormalized $W$ boson mass $m_W^{\textrm{reno}}$.
Their formulae are given in Appendix~\ref{sec:STU}.

 \subsection{Renormalization in the fermion sector}\label{sec:reno fermi}

We here discuss renormalization in the fermion sector.
The Lagrangian of the fermion sector is given by 
 \begin{align}
 \mathcal{L}_f = \bar{\Psi}_L i \slashed{\partial} \Psi_L 
             + \bar{\Psi}_R i \slashed{\partial} \Psi_R
             - m_f^{}(\bar{\Psi}_L \Psi_R + \bar{\Psi}_R \Psi_L),
 \end{align} 
where $\Psi_L^{}$ ($\Phi_R^{}$) is a left (right) handed fermion field. 
They are shifted into a renormalized parameter and renormalized fields, and counter terms,
 \begin{align}
      m_f^{} & \rightarrow m_f^{} + \delta m_f^{} , \\
 \Psi_L & \rightarrow \Psi_L + \frac{1}{2} \delta Z_L^f, \\ 
 \Psi_R & \rightarrow \Psi_R + \frac{1}{2} \delta Z_R^f.
 \end{align}

Two point functions of fermion fields are composed of following two parts,
 \begin{align}
 \hat{\Pi}_{ff}[p^2] = \hat{\Pi}_{ff,V}[p^2] + \hat{\Pi}_{ff,A}[p^2].
 \end{align} 
Each part is expressed,
 \begin{align}
 & \hat{\Pi}_{ff,V}[p^2] = 
  \slashed{p}\Pi_{ff,V}^{1\PI}[p^2] + \slashed{p}\delta Z_V^f 
  +m_f^{} \Pi_{ff,S}^{1\PI}[p^2] - m_f^{}\delta Z_V^f - \delta m_f^{} ,\\
 & \hat{\Pi}_{ff,A}[p^2] = 
  - \slashed{p} \gamma_5 \left(\Pi_{ff,A}^{1\PI}[p^2] + \delta Z_A^f \right),
 \end{align}
where
 \begin{align}
 \delta Z_V^f = \frac{1}{2}(\delta Z_L^f + \delta Z_R^f), \,\,\,\,\,
 \delta Z_A^f = \frac{1}{2}(\delta Z_L^f - \delta Z_R^f).
 \end{align}
We determine the counter terms by following conditions,
 \begin{align}
 \hat{\Pi}_{ff,V}[m_f^2] = 0, \,\,\,\,\,
 \frac{d}{d\slashed{p}}\hat{\Pi}_{ff,V}[p^2]\Big|_{p^2=m_f^2} = 0, \,\,\,\,\,
 \frac{d}{d\slashed{p}}\hat{\Pi}_{ff,A}[p^2]\Big|_{p^2=m_f^2} = 0.
 \end{align}
Then we obtain each counter term,
 \begin{align}
& \delta m_f^{} = m_f^{}\left(\Pi_{ff,V}^{1\PI}[m_f^2]+ \Pi_{ff,S}^{1\PI}[m_f^2]\right), \\
& \delta Z_V^f = -\Pi_{ff,V}^{1\PI}[m_f^2] 
             - 2m_f^2\left(\frac{d}{dp^2}\Pi_{ff,V}^{1\PI}[p^2]\Big|_{p^2=m_f^2}
                       + \frac{d}{dp^2}\Pi_{ff,S}^{1\PI}[p^2]\Big|_{p^2=m_f^2}\right), \\
& \delta Z_A = -\Pi_{ff,A}^{1\PI}[m_f^2] + 2m_f^2 \frac{d}{dp^2} \Pi_{ff,A}^{1\PI}[p^2]
 \Big|_{p^2=m_f^2}.
 \end{align} 

 \subsection{Renormalization in the Higgs sector}\label{sec:reno Higgs}

There are eight following parameters in the Higgs potential,
 \begin{align}
 m_\Phi^2, \lambda, \mu_{\Phi S}, \lambda_{\Phi S}, t_S, m_S^2, \mu_S, \lambda_S.
 \end{align}
As described in Sec.~\ref{sec2}, four of them can be rewritten in terms of the physical parameters $m_h^2$, $m_H^2$, $\alpha$ and $v$
by using Eqs.~(\ref{eq:mphi}), (\ref{eq:lambda}), (\ref{eq:ms}), (\ref{eq:mups}).
Remained parameters are $\lambda_{\Phi S}, v_S, \mu_S, \lambda_S$, 
where $t_S^{}$ is replaces by $v_S^{}$ as described in Eq.~(\ref{tad pole}).

First, we shift the bare parameters into renormalized parameters,
 \begin{align}
 & m_h^2 \rightarrow m_h^2 + \delta m_h^2, \\
 & m_H^2 \rightarrow m_H^2 + \delta m_H^2, \\
 & \alpha \rightarrow \alpha + \delta \alpha, \\
 & v \rightarrow v + \delta v, \\
 & \lambda_{\Phi S} \rightarrow \lambda_{\Phi S} + \delta \lambda_{\Phi S}, \\
 & v_S \rightarrow v_S + \delta v_S, \\
 & \lambda_S \rightarrow \lambda_S +\delta \lambda_S,\\
 & \mu_S \rightarrow \mu_S +\delta \mu_S.
 \end{align} 
Two physical scalar fields are shifted to the renormalized fields and the wave function renormalizations, 
 \begin{align}
 \left(\begin{array}{c}
       H \\
       h \\ 
       \end{array}\right) \rightarrow
 \left(\begin{array}{cc}
       1 + \frac{1}{2}\delta Z_h & \delta C_{hH} + \delta \alpha \\
       \delta C_{Hh} - \delta \alpha & 1 + \frac{1}{2} \delta Z_H \\
       \end{array} \right)
 \left(\begin{array}{c}
       H \\
       h \\
       \end{array}\right).
 \end{align}
We also shift the tadpoles as
 \begin{align}
 T_h \rightarrow T_h + \delta T_h, \,\,\, 
 T_H \rightarrow T_H + \delta T_H,
 \end{align}
where $T_H$ and $T_h$ are related with $T_\phi$ and $T_s$ as 
 \begin{align}
 \left(\begin{array}{c}
       T_s \\
       T_\phi \\
       \end{array}\right) = \left(\begin{array}{cc}
                                  c_\alpha^{} & -s_\alpha^{} \\
                                  s_\alpha^{} & c_\alpha^{} \\
                                  \end{array}\right)
 \left(\begin{array}{c}
       T_H \\
       T_h \\
       \end{array}\right).
 \end{align}

Renormalized one and two point functions at the one-loop level are given by
 \begin{align}
 \hat{\Gamma}_h &= 0 + \delta T_h + \Gamma_h^{1\PI}, \\
 \hat{\Gamma}_H &= 0 + \delta T_H + \Gamma_H^{1\PI}, \\
  \hat{\Pi}_{hh}[p^2] &=
       (p^2-m_h^2)(1+\delta Z_h) -\delta m_h^2 
    +\frac{c_\alpha^2}{v} \delta T_\phi 
    +\Pi_{hh}^{1\textrm{PI}}[p^2] , \\
 \hat{\Pi}_{hH}[p^2] &= (p^2-m_h^2)\delta C_{Hh} + (p^2- m_H^2)\delta C_{hH}
                 +(m_h^2 -m_H^2)\delta \alpha 
                 +\frac{c_\alpha s_\alpha}{v}\delta T_\phi 
                 +\Pi_{hH}^{1\textrm{PI}}[p^2] , \\
 \hat{\Pi}_{HH}[p^2] &= (p^2-m_H^2)(1+\delta Z_H) -\delta m_H^2
                 +\frac{s_\alpha^2}{v}\delta T_\phi +\Pi_{HH}^{1\textrm{PI}}[p^2],
 \end{align}
where analytic expressions of 1PI diagram parts are given in the Appendix~\ref{sec:1PI}.

We note that
there are 14 independent counter terms in the Higgs sector.
By imposing following nine renormalized on-shell conditions, 
 \begin{align}
  &\hat{\Gamma}_h = 0, \hspace{5mm} 
   \hat{\Gamma}_H = 0 , \\
  &  \hat{\Pi}_{hh}[m_h^2] = 0, \hspace{5mm}
    \hat{\Pi}_{HH}[m_h^2] = 0 ,  \\
 & \hat{\Pi}_{hH}[m_h^2] =0, \hspace{5mm}   
  \hat{\Pi}_{hH}[m_H^2] =0, \hspace{5mm}
  \delta C_{hH} = \delta C_{Hh} \equiv \delta C_h \\
 & \frac{d}{dp^2}\hat{\Pi}_{hh}[p^2]\Big|_{p^2=m_h^2} = 1,\hspace{5mm}
  \frac{d}{dp^2}\hat{\Pi}_{HH}[p^2]\Big|_{p^2=m_H^2} = 1.
 \end{align}
we determine following nine counter terms,   
 \begin{align}
  &  \delta T_h = -\Gamma_h^{1\textrm{PI}}, \hspace{5mm}
     \delta T_H = -\Gamma_H^{1\textrm{PI}}, \\
  &  \delta m_h^2 = \frac{c_\alpha^2}{v} \delta T_\phi 
                    + \Pi_{hh}^{1\textrm{PI}}[m_h^2], \hspace{5mm}
     \delta m_H^2 =  \frac{s_\alpha^2}{v} \delta T_\phi 
                    + \Pi_{HH}^{1\textrm{PI}}[m_H^2], \\
 &  \delta C_{hH}=\delta C_{Hh} (\equiv \delta C_h)= \frac{1}{2(m_H^2 - m_h^2)} 
   \left[\Pi_{hH}^{1\textrm{PI}}[m_h^2] -\Pi_{hH}^{1\textrm{PI}}[m_H^2]\right],\\
 &  \delta \alpha  = \frac{1}{2(m_H^2-m_h^2)} \left[
         \frac{2s_\alpha c_\alpha}{v}\delta T_\phi
         +\Pi_{hH}^{1\textrm{PI}}[m_h^2] +\Pi_{hH}^{1\textrm{PI}}[m_H^2] \right],\\
 &   \delta Z_h = -\frac{d}{dp^2} \Pi_{hh}^{1\PI}[p^2]\Big|_{p^2=m_h^2}, 
      \hspace{5mm}
     \delta Z_H = -\frac{d}{dp^2} \Pi_{HH}^{1\PI}[p^2]\Big|_{p^2=m_H^2}.
 \end{align}
As shown in Sec.~\ref{sec:reno gauge},
$\delta v$ can be determined by the renormalization in the gauge sector. 
We note that forms of $\delta \lambda_{\Phi S},\delta v_S, \delta \lambda_S$ and 
$\delta \mu_S$ cannot be determine above conditions.
These do not appear in the one-loop calculation of the $hVV$ and $hf\bar{f}$ vertices. 
When one-loop corrections to the triple scalar couplings such as the $hhh$ coupling, 
these counter terms have to be determined by additional renormalization conditions as discussed in Ref.~\cite{KOSY,2HDM_full} 
in the context of the THDM. 
The study of one-loop corrections to the triple Higgs boson coupling in the HSM is discussed elsewhere~\cite{HSM_hhh}.

 \subsection{Renormalized Higgs couplings}
In this subsection, we give formulae of the renormalized Higgs boson couplings. 
They are composed of the tree level part, the counter term part and the 1PI diagram part. 
The renormalized $hVV$ and $hf\bar{f}$ couplings are expressed as
 \begin{align}
 \hat{\Gamma}_{hVV}[p_1^2,p_2^2,q^2] &= \hat{\Gamma}_{hVV}^1[p_1^2,p_2^2,q^2]g^{\mu \nu}
   +\hat{\Gamma}_{hVV}^2[p_1^2,p_2^2,q^2] \frac{p_1^\nu p_2^\mu}{m_V^2}
   +i\hat{\Gamma}_{hVV}^3[p_1^2,p_2^2,q^2]\epsilon^{\mu\nu\rho\sigma} \frac{p_{1\rho} p_{2\sigma}}{m_V^2}, \label{eq:hvv_form}\\[+5pt]
 \hat{\Gamma}_{hff}[p_1^2,p_2^2,q^2] &= 
  \hat{\Gamma}_{hff}^S[p_1^2,p_2^2,q^2] + \gamma_5 \hat{\Gamma}_{hff}^P[p_1^2,p_2^2,q^2]
 +\slashed{p}_1\hat{\Gamma}_{hff}^{V1}[p_1^2,p_2^2,q^2]
 + \slashed{p}_2\hat{\Gamma}_{hff}^{V2}[p_1^2,p_2^2,q^2] \notag\\[+5pt]
&+ \slashed{p}_1 \gamma_5 \hat{\Gamma}_{hff}^{A1}[p_1^2,p_2^2,q^2] 
 + \slashed{p}_2 \gamma_5 \hat{\Gamma}_{hff}^{A2}[p_1^2,p_2^2,q^2]
 + \slashed{p}_1 \slashed{p}_2 \hat{\Gamma}_{hff}^T[p_1^2,p_2^2,q^2]
 + \slashed{p}_1\slashed{p}_2 \gamma_5 \hat{\Gamma}_{hff}^{TP}[p_1^2,p_2^2,q^2].
 \label{eq:hff_form}
 \end{align}  
Each renormalized form factor is given by,
 \begin{align}
 \hat{\Gamma}_{hVV}^i[p_1^2,p_2^2,q^2]&=
  \frac{2m_V^2}{v}\kappa_V^i + \delta \Gamma_{hVV}^i
  + M_{hVV,i}^{1\PI}[p_1^2,p_2^2,q^2], 
  \,\,\,\,\, (i= 1,2,3)\\
 \hat{\Gamma}_{hff}^j[p_1^2,p_2^2,q^2]&=
  -\frac{m_f^{}}{v}\kappa_f^{} + \delta \Gamma_{hff}^j
  + F_{hff,j}^{1\PI}[p_1^2,p_2^2,q^2], 
  \,\,\,\,\, (j= S, P, V1, V2, A1, A2, T, TP),
 \end{align}
where the counter terms are expressed as
 \begin{align}
 \delta \Gamma_{hVV}^{1}&=
  \frac{2m_V^2}{v}c_\alpha^{} \left(
   \frac{\delta m_V^2}{m_V^2} -\frac{\delta v}{v} 
   +\frac{s_\alpha^{}}{c_\alpha^{}}\delta C_h +\delta Z_V 
  +\frac{1}{2}\delta Z_h 
  \right), \\
 \delta \Gamma_{hVV}^2 &= \delta \Gamma_{hVV}^3 = 0, \\
 \delta \Gamma_{hff}^S&=
  -\frac{m_f}{v}c_\alpha^{} \left(
   \frac{\delta m_f}{m_f} -\frac{\delta v}{v} 
   +\frac{s_\alpha^{}}{c_\alpha^{}}\delta C_h +\delta Z_f^V 
  +\frac{1}{2}\delta Z_h
  \right), \\
 \delta \Gamma_{hff}^P &= \delta \Gamma_{hff}^{V1} = \delta \Gamma_{hff}^{V2} =
 \delta \Gamma_{hff}^{A1} = \delta \Gamma_{hff}^{A2} = \delta \Gamma_{hff}^{T} = 
 \delta \Gamma_{hff}^{TP} = 0,
 \end{align}
where $\delta m_W^2$, $\delta Z_V$ and $\delta v$ are given in Sec.~\ref{sec:reno gauge} and $\delta m_f^{}$ and $\delta Z_f^{V}$ are given in Sec.~\ref{sec:reno fermi}.
Tree level scaling factors $\kappa_V^i$ and $\kappa_f^j$ are given by
 \begin{align}
 & \kappa_V^1 = \kappa_f^S =c_\alpha^{}, \\
 & \kappa_V^2 = \kappa_V^3 =0 ,\\
 & \kappa_V^P = \kappa_V^{V1} = \kappa_V^{V2} = \kappa_V^{A1} = \kappa_V^{A2} = 
   \kappa_V^{T} = \kappa_V^{TP} = 0.
 \end{align}

\section{Numerical evaluation for the scaling factors}

\subsection{Renormalized scaling factors} 
 In this section, we show numerical results for the renormalized Higgs boson couplings, i.e., $hVV$ and $hf\bar{f}$. 
We also calculate the leading order results of the decay rate of the $h$ to $\gamma\gamma$ process. 
 Our numerical program is written as a FORTRAN program, and the package; LoopTools~\cite{LoopTools} is used for the one-loop integrations. 

Our numerical results are shown in terms of the scaling factors.
Deviations in  the one-loop corrected scaling factors for $hVV$ and $hf\bar{f}$ couplings are defined as 
 \begin{align}
 \Delta \hat{\kappa}_V \equiv 
 \frac{\hat{\Gamma}_{hVV}^1[p_1^2,p_2^2,q^2]}
           {\hat{\Gamma}_{hVV,\SM}^1[p_1^2,p_2^2,q^2]} -1, \\
 \Delta \hat{\kappa}_f \equiv 
 \frac{\hat{\Gamma}_{hff}^S[p_1^2,p_2^2,q^2]}
           {\hat{\Gamma}_{hff,\SM}^S[p_1^2,p_2^2,q^2]} -1, 
 \end{align}
where $\hat{\Gamma}_{hVV,\SM}^1$ and $\hat{\Gamma}_{hff,\SM}^S$ are the one-loop corrected $hVV$ and $hf\bar{f}$ couplings in the SM. 
The formulae for the one-loop decay rates $h\rightarrow \gamma\gamma$, $h\rightarrow Z\gamma$ and $h \rightarrow gg$ are given in Appendix~\ref{sec:decay rate}.
We numerically evaluate deviations in the scaling factor of the $h\gamma\gamma$ , $h\gamma Z$ and $hgg$ effective coupling defined as
 \begin{align}
 \Delta \kappa_\gamma & \equiv \sqrt{\frac{\Gamma[h\rightarrow \gamma\gamma]}
           {\Gamma[h\rightarrow \gamma\gamma]_\SM}}  - 1, \\
 \Delta \kappa_{\gamma Z} & \equiv \sqrt{\frac{\Gamma[h\rightarrow \gamma Z]}
           {\Gamma[h\rightarrow \gamma Z]_\SM}}  - 1, \\
 \Delta \kappa_g & \equiv \sqrt{\frac{\Gamma[h\rightarrow gg]}
           {\Gamma[h\rightarrow gg]_\SM}}  - 1,
 \end{align}
where $\Gamma[h\rightarrow XY]$ ($\Gamma[h\rightarrow XY]_\SM$) is the prediction of the decay rate for $h\rightarrow XY$ mode in the HSM (in the SM). 
Because the additional Higgs boson does not have electromagnetic charge and color charge, decay rates ($\Gamma[h\rightarrow XY]$) of these modes are modified only by field mixing effects at the one-loop level.
The scaling factor of the $h\gamma\gamma$, $h\gamma Z$ and $hgg$ vertex, $\Delta \kappa_\gamma$, $\Delta \kappa_{\gamma Z}^{}$ and $\Delta \kappa_g^{}$ are given by 
 \begin{align}
 \Delta \kappa_\gamma^{} = \Delta \kappa_{\gamma Z}^{}
  = \Delta \kappa_g^{} = c_\alpha^{} - 1.
 \end{align}

In our numerical evaluation, we use the following values for the input parameters~\cite{PDG}:
 \begin{align}
 & G_F = 1.1663787 \time 10^{-5}\,\GeV^{-2},\,\,\,
   m_Z^{} = 91.1876\, \GeV, \,\,\,
   \alpha_{\EM}^{} = 1/137.035999074,\,\,\,
   \Delta \alpha_{\EM} = 0.06637, \notag\\
 & m_t^{} = 173.21\, \GeV, \,\,\, m_b^{} = 4.66\,\GeV,\,\,\,
   m_c^{} = 1.275\,\GeV,\,\,\, m_\tau^{} = 1.77682\,\GeV, \notag\\
 &  m_h^{} = 125\GeV, 
 \end{align}
where $\Delta\alpha_{\EM}^{}$ is 
defined as $1-\frac{\alpha_{\EM}^{}}{\hat{\alpha}_{\EM}^{}(m_Z^{})}$ with $\hat{\alpha}_{\text{em}}(m_Z)$ being the fine structure constant at the scale of $m_Z$.
Furthermore, we set the momenta $(p_1^2, p_2^2,q^2)$ to be $(m_V^2, m_h^2, (m_h+m_V)^2)$ and $(m_f^2, m_f^2, m_h^2)$ for 
$\Delta \hat{\kappa}_V^{}$ and $\Delta \hat{\kappa}_f^{}$, respectively.
As we mentioned in Sec.~\ref{sec2}, 
we can take the value of $v_S^{}$ freely without changing physics.
We fix $v_S^{}$ to be 0 in the following numerical analyses.

\subsection{One-loop corrections to the scaling factors in the HSM}

First, we discuss approximate formulae in the case for $\alpha =0$ 
which can be expressed following simple forms
 \begin{align}
 & \Delta \hat{\kappa}_Z^{} = \Delta \hat{\kappa}_f ^{} \simeq -\frac{1}{16\pi^2}\frac{1}{6}\frac{m_H^2}{v^2}\left(1-\frac{\mathcal{M}^2}{m_H^2}\right)^2. \label{eq:dkz_0}
 \end{align}
The most right hand side of Eq.~(\ref{eq:dkz_0}) comes from the $H$ loop contributions of $\delta Z_h^{}$.
The structures of these one-loop contributions are the same as those in the THDMs as described in Ref.~\cite{2HDM_full}. 
As we can see in Eq.~(\ref{eq:dkz_0}) that there appears the quadratic mass like dependence in the one-loop correction to $hVV$ and $hf\bar{f}$ couplings when $\mathcal{M}^2 \ll m_H^2$, 
which can be regarded as the non-decoupling $H$ loop effect. 
If the mass of $H$ is mainly given by $\mathcal{M}^2$, this non-decoupling effect vanishes due to the factor $(1-\mathcal{M}^2/m_H^2)^2$ in Eq.~(\ref{eq:dkz_0}).

 \begin{figure}[t]
 \centering
  \includegraphics[width=7cm]{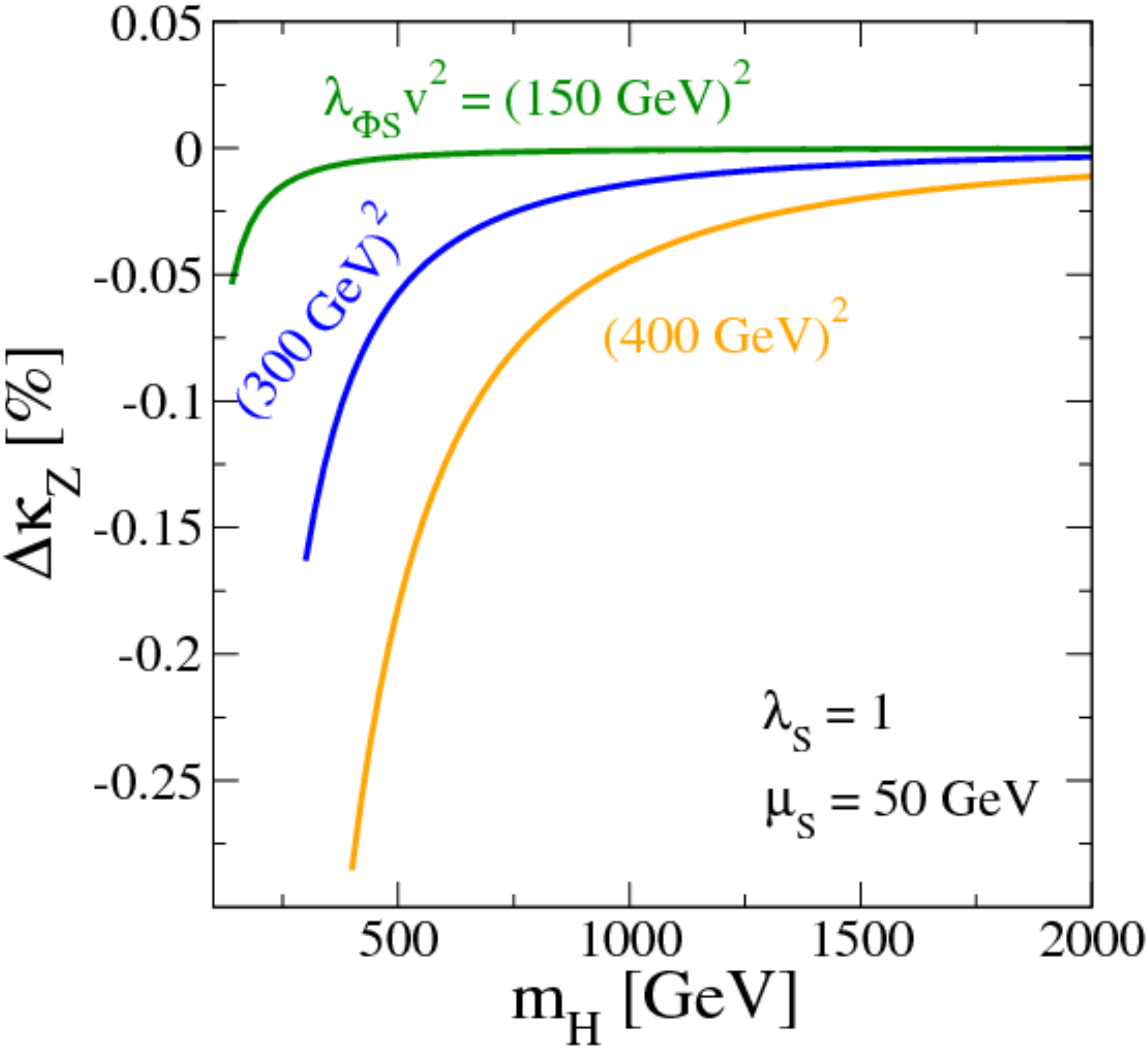} \hspace{0.5cm}
  \includegraphics[width=7cm]{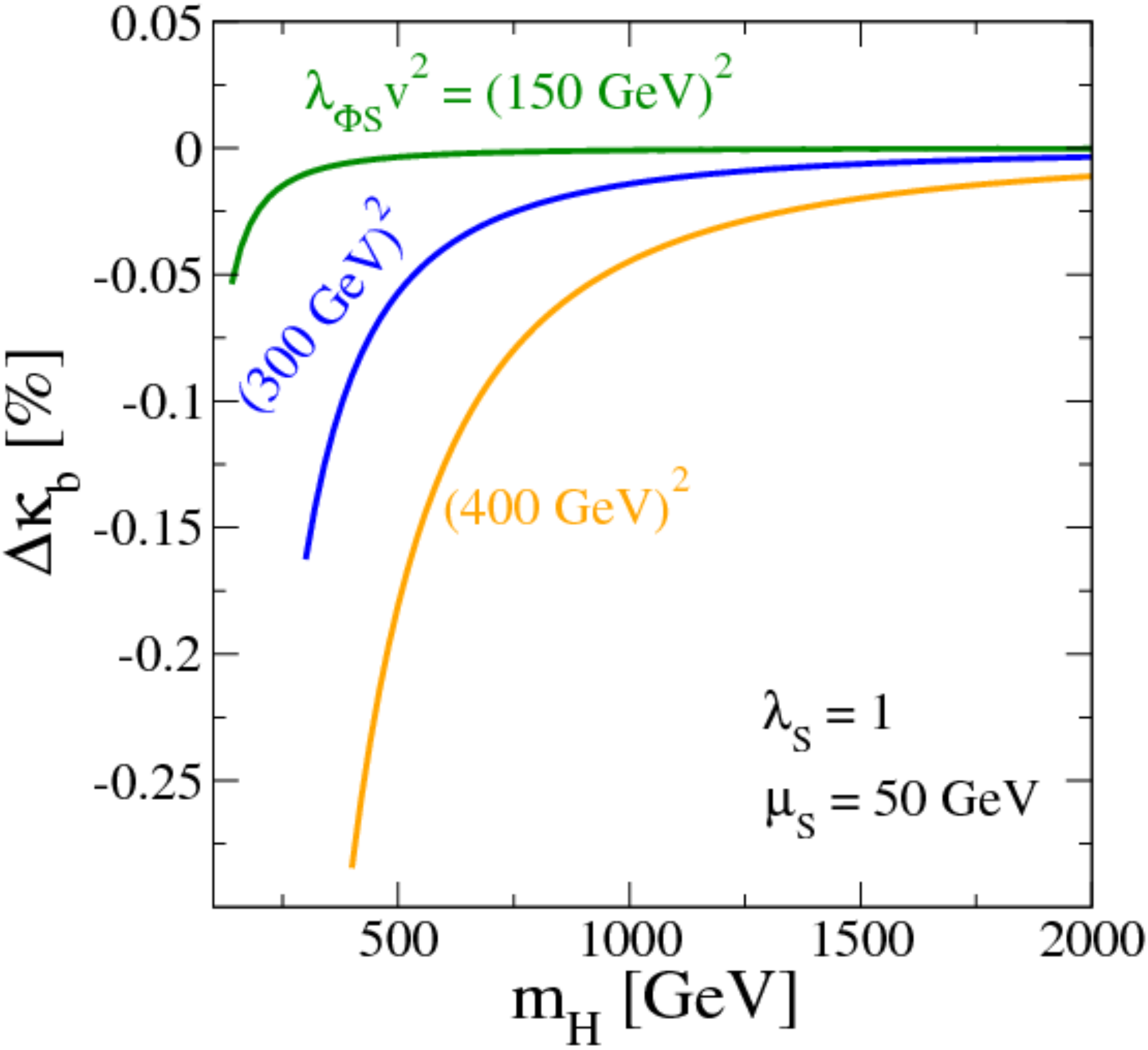} \hspace{0.5cm}
  \caption{$\Delta \hat{\kappa}_Z^{}$ (left panel) and $\Delta \hat{\kappa}_b^{}$ (right panel) as a function of $m_H^{}$ under the constraints from perturbative unitarity, vacuum stability and the condition to avoid the wrong vacuum in the case for $\alpha = 0$.
We take $\lambda_S^{} =1$, $\mu_S^{}=50$ GeV and $\mathcal{M}^2>0$.
Green, blue and orange curves are the results for $\lambda_{\Phi S} v^2 = (150\GeV)^2$, $(300\GeV)^2$ and $(400\GeV)^2$, respectively.}
  \label{decouple}
 \end{figure}

 \begin{figure}[t]
 \centering
  \vspace{0.5mm}
  \includegraphics[width=7cm]{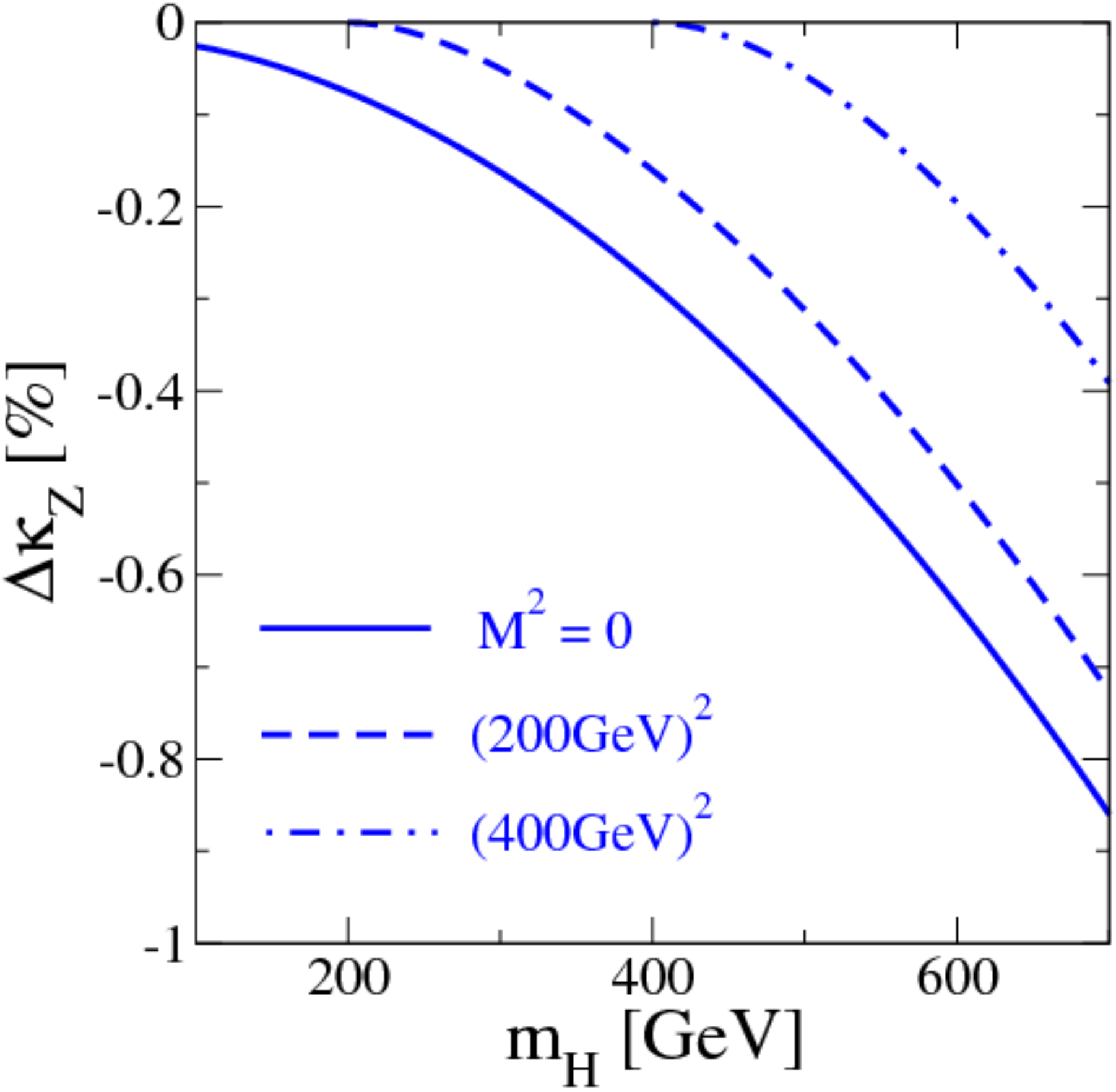} \hspace{0.5cm}
  \includegraphics[width=7cm]{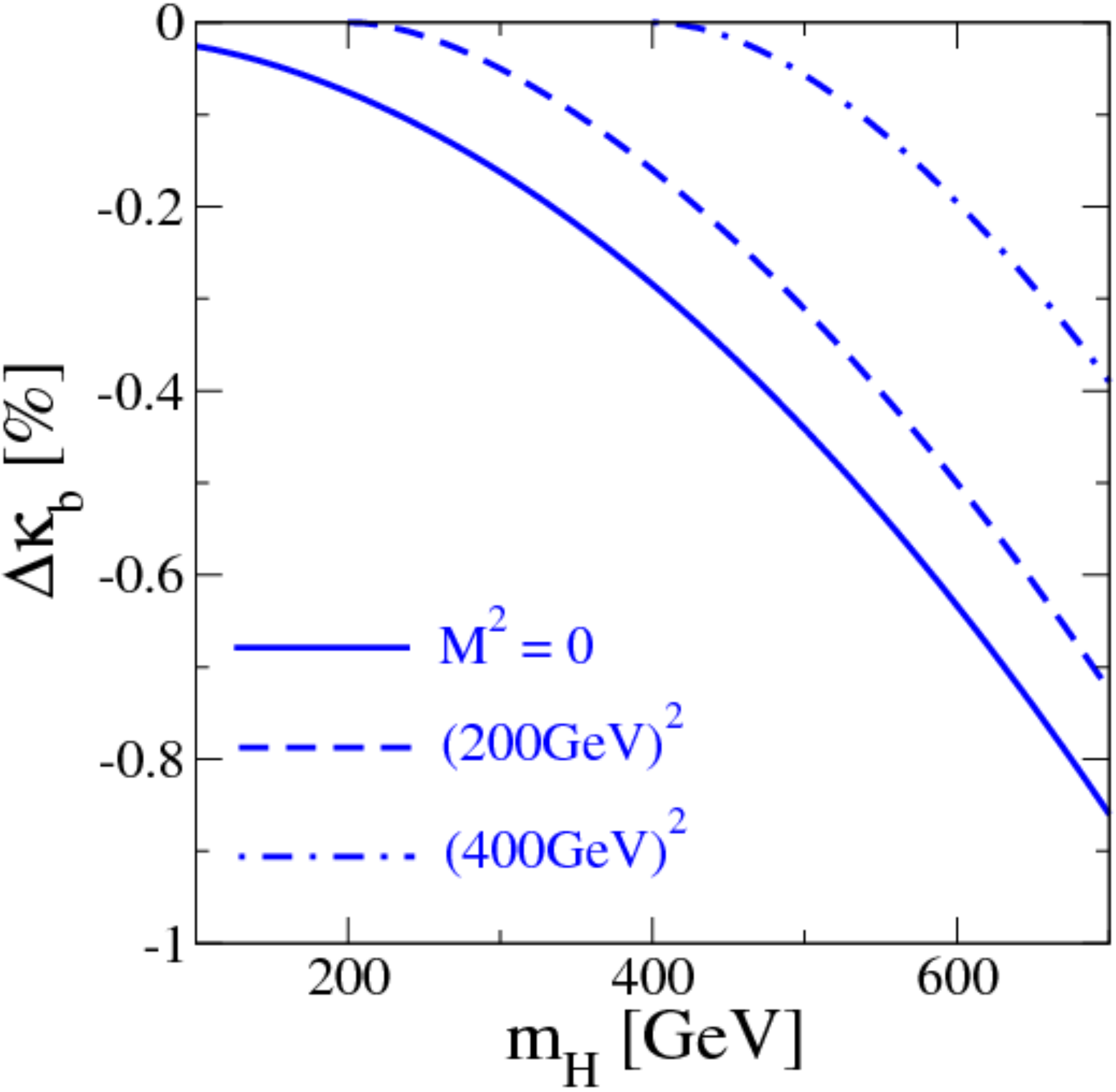} \hspace{0.5cm}
  \caption{$\Delta \hat{\kappa}_Z^{}$ (left panel) and $\Delta \hat{\kappa}_b^{}$ (right panel) as a function of $m_H^{}$ under the constraints from perturbative unitarity, vacuum stability and the condition to avoid the wrong vacuum in the case for $\alpha = 0$.
We take $\lambda_S^{} =1$, $\mu_S^{}=50$ GeV and $\lambda_{\Phi S}^{}>0$.
Solid, dashed and dot-dash curves are the results for  $\mathcal{M}^2 = 0$, $(200\GeV)^2$ and $(400\GeV)^2$, respectively.}
  \label{nonde}
 \end{figure}

In Fig.~\ref{decouple}, we show the decoupling behavior of $H$ loop contributions to the renormalized Higgs boson couplings under the constraints from perturbative unitarity, vacuum stability and the condition to avoid the wrong vacuum in the case for $\alpha = 0$. 
The left and right panels are $\Delta \hat{\kappa}_Z^{}$ and $\Delta \hat{\kappa}_b^{}$ as a function of $m_H^{}$, respectively. 
We fix $\lambda_S^{} = 1$ and $\mu_S^{}=50$ GeV.
Green, blue and orange curves indicate predictions for $\lambda_{\Phi S} v^2 = (150\GeV)^2$, $(300\GeV)^2$ and $(400\GeV)^2$, respectively. 
Since the value of $\mathcal{M}^2$ grows as $m_H^2$ becomes large, 
we can see that deviations by loop effects are reduced in the large mass regions.

In Fig.~\ref{nonde}, we show $\Delta \hat{\kappa}_Z^{}$ (the left panel) and $\Delta \hat{\kappa}_b^{}$ (the right panel) as a function of $m_H^{}$ 
in the case for $\alpha=0$.
We fix $\lambda_S^{}=1$ and $\mu_S^{}=50$ GeV. 
We investigate the behavior of $\Delta\hat{\kappa}_X^{}$ for various values of $\mathcal{M}^2$ such as $\mathcal{M}^2 = 0$, $(200 \GeV)^2$ and $(400\GeV)^2$.
In this case, the magnitude of the deviations increase when $m_H^{}$ becomes large in each the Higgs coupling because of the non-decoupling effect of $m_H^{}$.

 \begin{figure}[t]
 \centering
  \includegraphics[width=7cm]{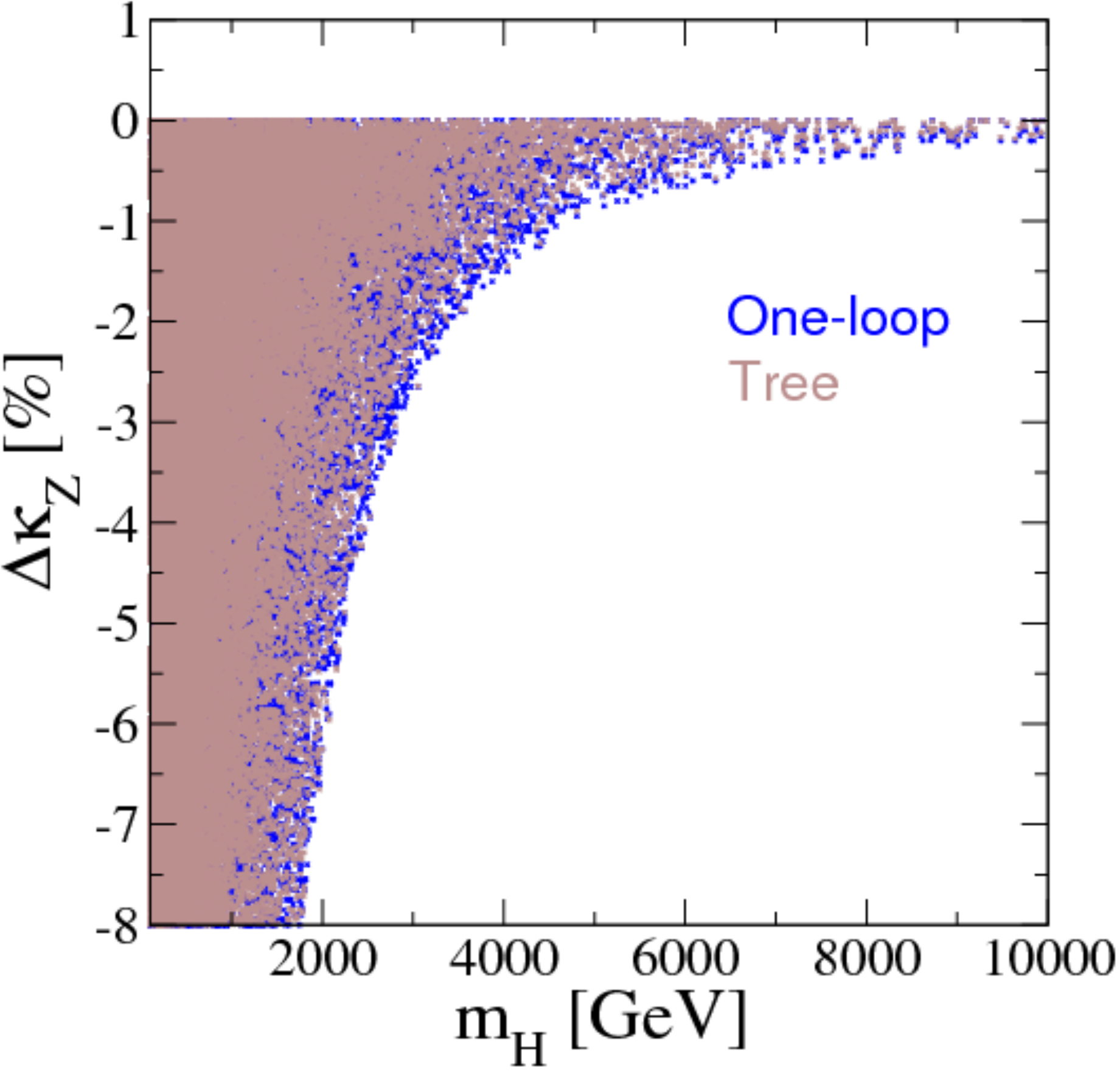} \hspace{0.5cm}
  \includegraphics[width=7cm]{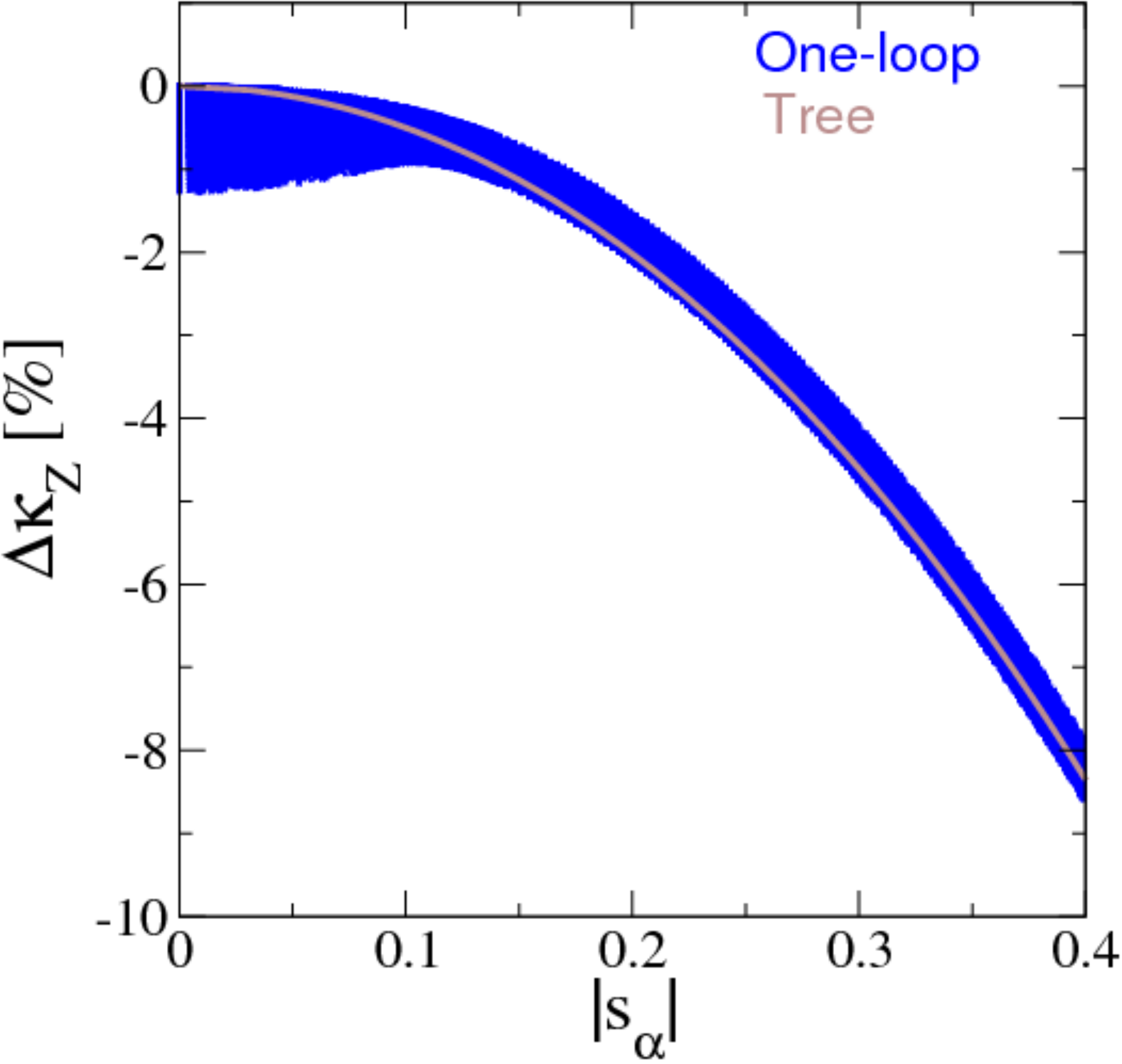} \hspace{0.5cm}
  \caption{Scatter plots of allowed regions under the constraints of perturbative unitarity, vacuum stability and the condition of the wrong vacuum 
on the $m_H^{}$-$\Delta\kappa_Z^{}$ plane (left panel) and the $|s_\alpha^{}|$-$\Delta\kappa_Z^{}$ plane (right panel). 
Brown points are the results of the tree level calculation, while blue points are those of the one-loop calculation.
Parameters are scanned as $100$ GeV $< m_H^{} < 10$ TeV, $0.91 \leq c_\alpha^{} \leq 1.00$ and $-m_H^2 < \mathcal{M}^2 < m_H^2$ with fixing 
$\lambda_S^{}=0.1$ and $\mu_S^{}=0$.}
  \label{non_zero}
 \end{figure}

In Fig.~\ref{non_zero}, we show scatter plots of allowed regions under the constraints of perturbative unitarity, vacuum stability and the condition of the wrong vacuum 
on the $m_H^{}$-$\Delta\kappa_Z^{}$ plane (left panel) and the $|s_\alpha^{}|$-$\Delta\kappa_Z^{}$ plane (right panel). 
Brown points are the results of the tree level calculation, while blue points are those of the one-loop calculation.
We scan parameters as $100$ GeV $< m_H^{} < 10$ TeV, $0.91 \leq c_\alpha^{} \leq 1.00$ and $-m_H^2 < \mathcal{M}^2 < m_H^2$ with fixing 
$\lambda_S^{}=0.1$ and $\mu_S^{}=0$.
In Fig.~\ref{non_zero} (left), we learn that $\Delta\hat{\kappa}_Z^{}$ is zero in the large mass limit for $H$. 
For a nonzero negative value of $\Delta\kappa_Z^{}$ there is an upper bound on $m_H^{}$. 
The upper bound evaluated at the one-loop level is almost the same as that at the tree level for each value of negative $\Delta \kappa_Z^{}$. 
If by future precision measurements $\Delta\kappa_Z^{}$ is determined as $\Delta\kappa_Z^{}= -2.0 \pm 0.5\%$, 
the upper bound on $m_H^{}$ is obtained to be about 4 TeV. 
In Fig.~\ref{non_zero} (right), the tree level results are on the curve described by $\sim - s^2_\alpha/2$ for small $|s_\alpha^{}|$. 
At the one-loop level the magnitude of the deviation from the tree level prediction is typically about 1\%. 
For smaller values of $|s_\alpha^{}|$, $\Delta\hat{\kappa}_Z^{}$ is smaller than the tree level prediction, 
while for larger  $|s_\alpha^{}|$ the one-loop corrected value $\Delta\hat{\kappa}_Z^{}$ can be larger than the tree level prediction 
but the sign of $\Delta\hat{\kappa}_Z^{}$ is always negative.   

We omit the scan analysis of $\Delta \hat{\kappa}_f^{}$ in case for $\alpha \neq 0$, 
because results of $\Delta \hat{\kappa}_f^{}$ are almost the same as those of $\Delta \hat{\kappa}_V^{}$ as shown in Eq.~(\ref{eq:dkz_0}).

\section{Fingerprinting HSM and THDM by using future precision measurements}

In this section, we demonstrate how we can distinguish the simplest extended Higgs sectors by using one-loop corrected Higgs couplings and 
future precision measurements of the Higgs boson couplings.
In Ref.~\cite{KTYY}, the patterns of deviations in these couplings have been discussed at the tree level in the extended Higgs sectors which predict $\rho = 1$ 
at the tree level; i.e.,   
four types of THDMs with the softly broken discrete $Z_2$ symmetry~\cite{four_type_2HDMs, Type_X}, 
the HSM~\cite{HSM_pheno}, 
the Georgi-Machacek (GM) model where additional real and complex triplet scalar fields are introduced~\cite{GM}, and the model with the septet scalar field~\cite{exotic,septet}.
It has been shown that four types of THDMs (Type I, Type II, Type X and Type Y) can be basically separated by measuring Yukawa coupling constants of 
$h\tau\tau$, $hb\bar{b}$, $hc\bar{c}$ and/or $ht\bar{t}$ except for the decoupling regions.
On the other hand, the Type I THDM, in which only one of the Higgs doublets couples to all the fermions, 
and all the other extended Higgs sectors (the HSM, the GM model and the model with a septet field) can be distinguished by the precision measurement of the $hVV$ coupling and the universal coupling of $hf\bar{f}$ as long as the deviations in $\kappa_V^{}$ is detected.   
One of the notable features of the predictions in the {\it exotic} extended Higgs sectors such as the GM model and the model with the septet field is the prediction that the scaling factor $\kappa_V^{}$ 
can be greater than unity~\cite{GM, exotic,septet, HTM_models}, 
while both THDMs and the HSM always predict $\kappa_V^{} \leq 1$.

In order to compare the theory calculations with precision measurements at future lepton colliders such as the ILC, 
where most of the Higgs couplings are expected to be measured with high accuracies at the typically $\mathcal{O}(1)$ \% level or even better~\cite{HWGR},
the above tree level analyses in Ref.~\cite{KTYY} must be improved by using the predictions with radiative corrections.
In Refs.~\cite{2HDM_yukawa_reno, 2HDM_full}, the one-loop corrected scaling factors in the four types of THDMs have been calculated in the on-shell scheme, and the above tree level discussions 
in Ref.~\cite{KTYY} have been repeated but at the one-loop level. 
Even in the case including one-loop corrections, 
it is useful to discriminate types of Yukawa interactions 
by using the pattern of deviations among the $hf\bar{f}$ couplings. 
It is also demonstrated in Ref.~\cite{2HDM_full} that information of inner parameters can be considerably extracted by combination of the precision measurements 
on the Higgs boson couplings when a deviation in $\kappa_V^{}$ is large enough to be detected.  

  \begin{table}[t]
  \renewcommand{\arraystretch}{1.5}
   \centering
   \caption{Range of the parameter scan in the HSM and the Type I THDM in Figs.~\ref{dkv_dkb}, Fig.~\ref{dkv_dkg} and Fig.~\ref{dkv_dkg2}. For the definition of the parameters in the THDM, see; e.g., Ref.~\cite{2HDM_full}}
   \begin{tabular}{|c|c|}
   \hline
   HSM & THDM  \\\hline\hline
   $300 \GeV < m_H^{} < 1 \textrm{TeV}$ &
   $300 \GeV < m_H^{}(=m_A^{}=m_{H^\pm}^{}) < 1 \textrm{TeV}$ 
   \\
   $c_\alpha^{}  < 1$ &
   $\sin(\beta-\alpha) < 1$ 
   \\
   $  -15  <  \lambda_{\Phi S} < 15  $ &
   $0 < M^2 < (1\textrm{TeV})^2 $ 
   \\
   $  -15 <  \lambda_{S}^{} < 15 $ & 
   \\
   $  -2 \textrm{TeV} <  \mu_S^{} < 2\textrm{TeV} $ &  
   \\\hline
   \end{tabular}
   \label{scan range}
  \end{table}

We here show the one-loop corrected scaling factors of $hZZ$ and $hb\bar{b}$ couplings in the HSM in comparison with those in the Type I THDM.
The expected 1$\sigma$ uncertainties for these scaling factors at the LHC with the center-of-mass energy ($\sqrt{s}$) to be 14 TeV 
and the integrated luminosity ($L$) to be 3000 fb$^{-1}$ (HL-LHC) 
and also the ILC with the combination of the run with $\sqrt{s}=250$ GeV with $L = 250$ fb$^{-1}$ and that with $\sqrt{s}=500$ GeV with $L = 500$ fb$^{-1}$ (ILC500) are 
given by~\cite{HWGR}
 \begin{align}
 &  [\sigma(\kappa_Z),\sigma(\kappa_b), \sigma(\kappa_\gamma)] =
    [2\%, 4\%, 2\%], \hspace{1cm} \textrm{HL-LHC}, \notag\\
 &  [\sigma(\kappa_Z),\sigma(\kappa_b), \sigma(\kappa_\gamma)] =
    [0.49\%, 0.93\%, 8.3\%], \hspace{1cm} \textrm{ILC500}, 
 \end{align}
For the predictions at the one-loop level in the THDM, we fully use the formulae and the numerical program developed in Ref.~\cite{2HDM_full}. 

 \begin{figure}[tph]
 \centering
  \includegraphics[width=7cm]{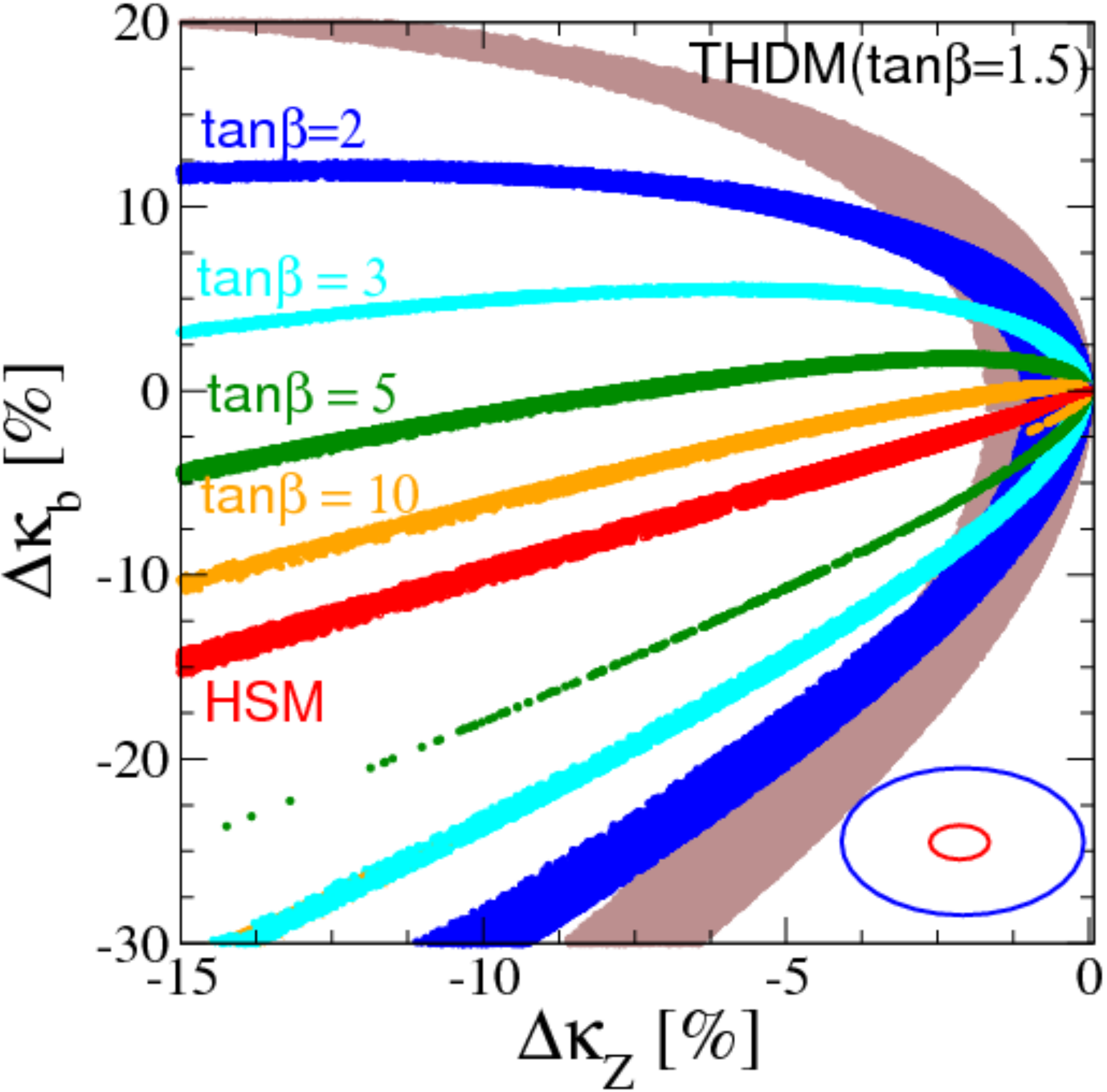} \\\vspace{0.5cm}
  \caption{Predictions of the allowed regions of the HSM and the Type I THDM at the one-loop level on the plane of $\Delta \kappa_Z^{}$ and
$\Delta \kappa_b^{}$.
The inner parameters are scanned under the constraints of perturbative unitarity, vacuum stability and the condition to avoid the wrong vacuum, 
 given in Appendix~\ref{sec:constraints}. 
The list of scanned parameters and scanned range of these parameters are shown in Tab.~\ref{scan range}.
Red regions indicate the predictions of the HSM.
Brown, blue, cyan, green and orange regions are predictions in the Type I THDM for $\tan\beta=1.5$, 2, 3, 5 and 10, respectively.
Blue and red ellipses show $\pm 1 \sigma$ uncertainty for measurements of $\Delta \kappa_Z^{}$ and $\Delta \kappa_b^{}$ at the HL-LHC and the ILC500, respectively~\cite{HWGR}.  }
  \label{dkv_dkb}
 \end{figure}

In Fig.~\ref{dkv_dkb}, we show the one-loop corrected predictions of the allowed regions of the HSM and the Type I THDM on the plane of $\Delta \hat{\kappa}_Z^{}$ and $\Delta \hat{\kappa}_b^{}$.
The inner parameters are scanned under the constraints of perturbative unitarity~\cite{Uni, THDM_uni2,THDM_uni3,THDM_uni4,THDM_uni5}, vacuum stability~\cite{Fuyuto, THDM_vs1,THDM_vs2,THDM_vs3} and the condition to avoid the wrong vacuum~\cite{chen_dawson_lewis,wrong_vacuum}. 
which are shown given in Appendix~\ref{sec:constraints}. 
The list of scanned parameters and scanned ranges of these parameters are shown in Tab.~\ref{scan range}.
Red regions indicate the predictions of the HSM.
Brown, blue, cyan, green and orange regions are the allowed regions in the Type I THDM for $\tan\beta=1.5$, 2, 3, 5 and 10, respectively, 
with varied $m_\Phi^{}$($m_H^{}=m_A^{}=m_{H^\pm}^{}$) and $M$, 
where definitions of the parameters are given in Ref.~\cite{2HDM_full}.
The blue and red ellipses show the measurement uncertainties ($\pm 1 \sigma$) for $\Delta \kappa_Z^{}$ and $\Delta \kappa_b^{}$ at the HL-LHC and the ILC500~\cite{HWGR}, respectively.

First, we discuss the behavior for predictions of the HSM in Fig.~\ref{dkv_dkb}.
We find that magnitude of the deviations in the one-loop corrected $\kappa_Z^{}$ and $\kappa_b^{}$ are almost similar in the HSM. 
The reason is that tree level predictions of $\Delta \kappa_Z^{}$ and $\Delta \kappa_b^{}$ take a common form ($c_\alpha^{} - 1$). 
Namely, $\Delta \hat{\kappa}_Z^{}$ and $\Delta \hat{\kappa}_b^{}$ dominantly deviate from the SM predictions to the directions with the rate $1:1$ 
by the mixing effect, and the small width of the line of $1:1$ is made by the one-loop contributions.

Next, we explain the behavior for predictions of the Type I THDM. 
The scaling factors for the $hVV$ couplings at the tree level are different from those of $hf\bar{f}$ couplings~\cite{KTYY}.
In the case for $\cos(\beta-\alpha) < 0$, $\Delta \hat{\kappa}_b$ in the THDM ($\Delta \hat{\kappa}_b^\textrm{THDM}$) is negative and its magnitude is greater than $\Delta \hat{\kappa}_b^{}$ in the HSM ($\Delta \hat{\kappa}_b^\textrm{HSM}$) for the same deviation in $\Delta \hat{\kappa}_Z^{}$. 
On the other hand, in the case of $\cos(\beta- \alpha)>0$, $\Delta \hat{\kappa}_b^\textrm{THDM}$ is larger than $\Delta \hat{\kappa}_b^\textrm{HSM}$ for the same value of $\Delta \hat{\kappa}_Z^{}$. 
$\Delta \hat{\kappa}_Z^{}$ and $\Delta \hat{\kappa}_b^{}$ deviate from the SM predictions according to the tree level predictions due to the tree level mixing effect, 
and the one-loop contributions make the deviations from the tree level prediction by typically a few \%.
We find that $\Delta \hat{\kappa}_Z^{}$ and $\Delta \hat{\kappa}_b^{}$ are substantially modified by radiative corrections 
in the case for low $\tan\beta$ values than the case for large $\tan\beta$ values.
As the value of $\tan\beta$ become large, 
the $\Delta \hat{\kappa}_Z^{}$ and $\Delta \hat{\kappa}_b^{}$ plane prediction in the Type I THDM approximate the line of $1:1$.
Larger deviations in $\Delta\hat{\kappa}_Z^{}$ and $\Delta \hat{\kappa}_b^{}$ in the case with $\tan\beta \geq 10$, $\cos(\beta-\alpha)<0$ and $m_{\Phi}^{} > 300$ GeV are excluded by the constraints from perturbative unitarity and vacuum stability. 

 \begin{figure}[tph]
 \centering
  \includegraphics[width=7cm]{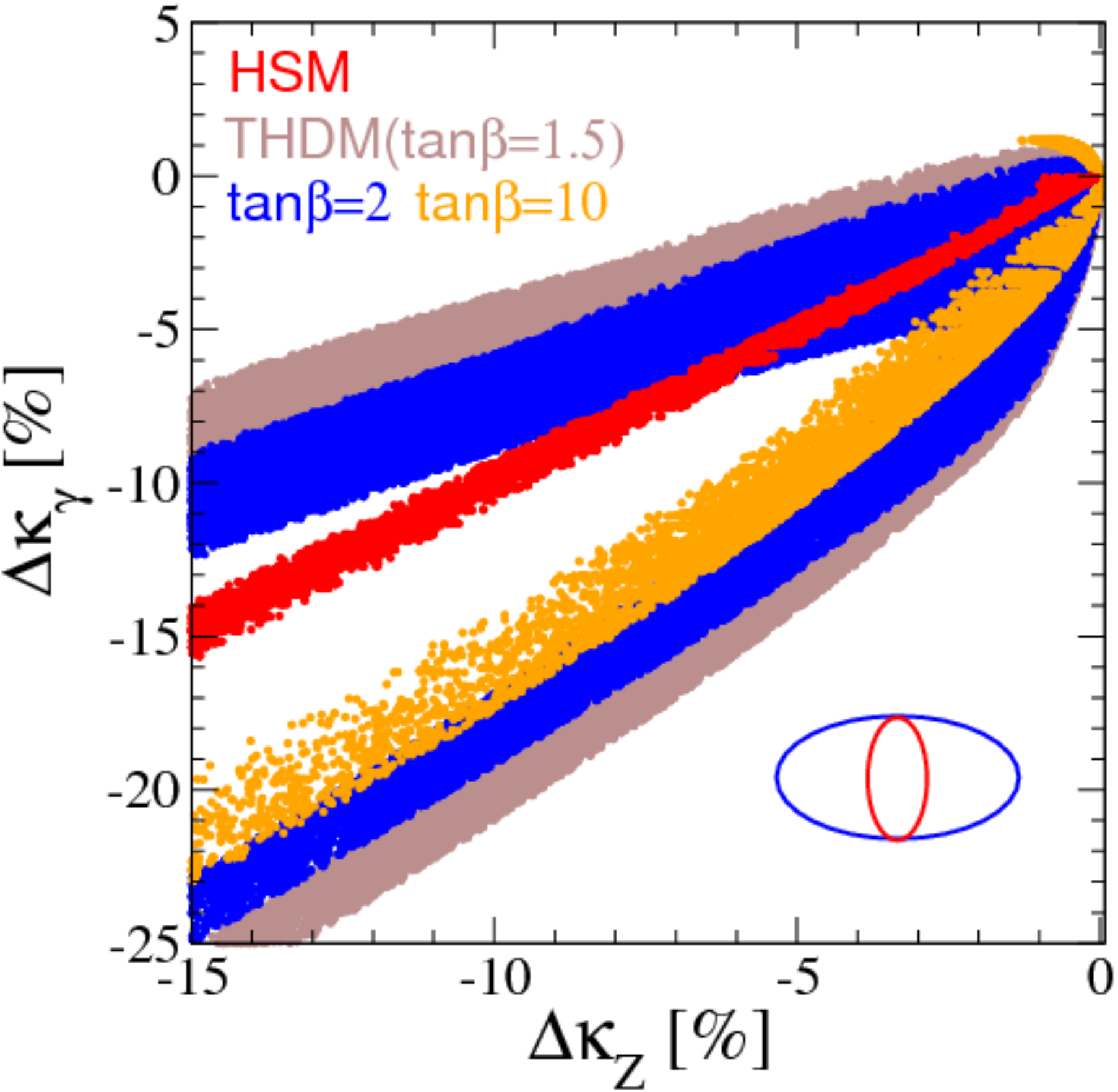} \vspace{0.5cm}
  \caption{Predictions of the allowed regions of the HSM and the Type I THDM on the plane of one-loop corrected $\Delta \kappa_Z^{}$ and
$\Delta \kappa_\gamma^{}$.
Blue ellipse is shown measurement uncertainties ($\pm 1 \sigma$) for $\Delta \kappa_Z^{}$ and $\Delta \kappa_\gamma^{}$ at the HL-LHC~\cite{HWGR}.
Red one is shown measurement uncertainties ($\pm 1 \sigma$) for $\Delta \kappa_Z^{}$ at the ILC500~\cite{HWGR} and $\Delta \kappa_\gamma^{}$ at the HL-LHC.
The others are same as in Fig.~\ref{dkv_dkb}.}
  \label{dkv_dkg}
 \end{figure}

In Fig.~\ref{dkv_dkg}, we show the one-loop corrected predictions of the allowed regions of the HSM and the Type I THDM on the plane of $\Delta \hat{\kappa}_Z^{}$ and
$\Delta \kappa_\gamma^{}$.
We scan inner parameters in each model within the ranges listed in Tab.~\ref{scan range} under the constraints of perturbative unitarity, vacuum stability and the condition to avoid the wrong vacuum 
and the condition to avoid the wrong vacuum. 
Definitions of color for allowed regions are the same as those in Fig.~\ref{dkv_dkb}. 
Blue and red ellipses are shown measurement uncertainties ($\pm 1 \sigma$) for $\Delta \hat{\kappa}_Z^{}$ and $\Delta \kappa_\gamma^{}$ at the HL-LHC and the ILC500~\cite{HWGR}.  
Since uncertainty of $\Delta \kappa_\gamma$ measurement at the HL-LHC is smaller than that at the ILC500,
we here use expected uncertainty for $\Delta \kappa_\gamma^{}$ at the HL-LHC in both case the HL-LHC and the ILC500.

\begin{figure}[tph]
 \centering
  \includegraphics[width=6cm]{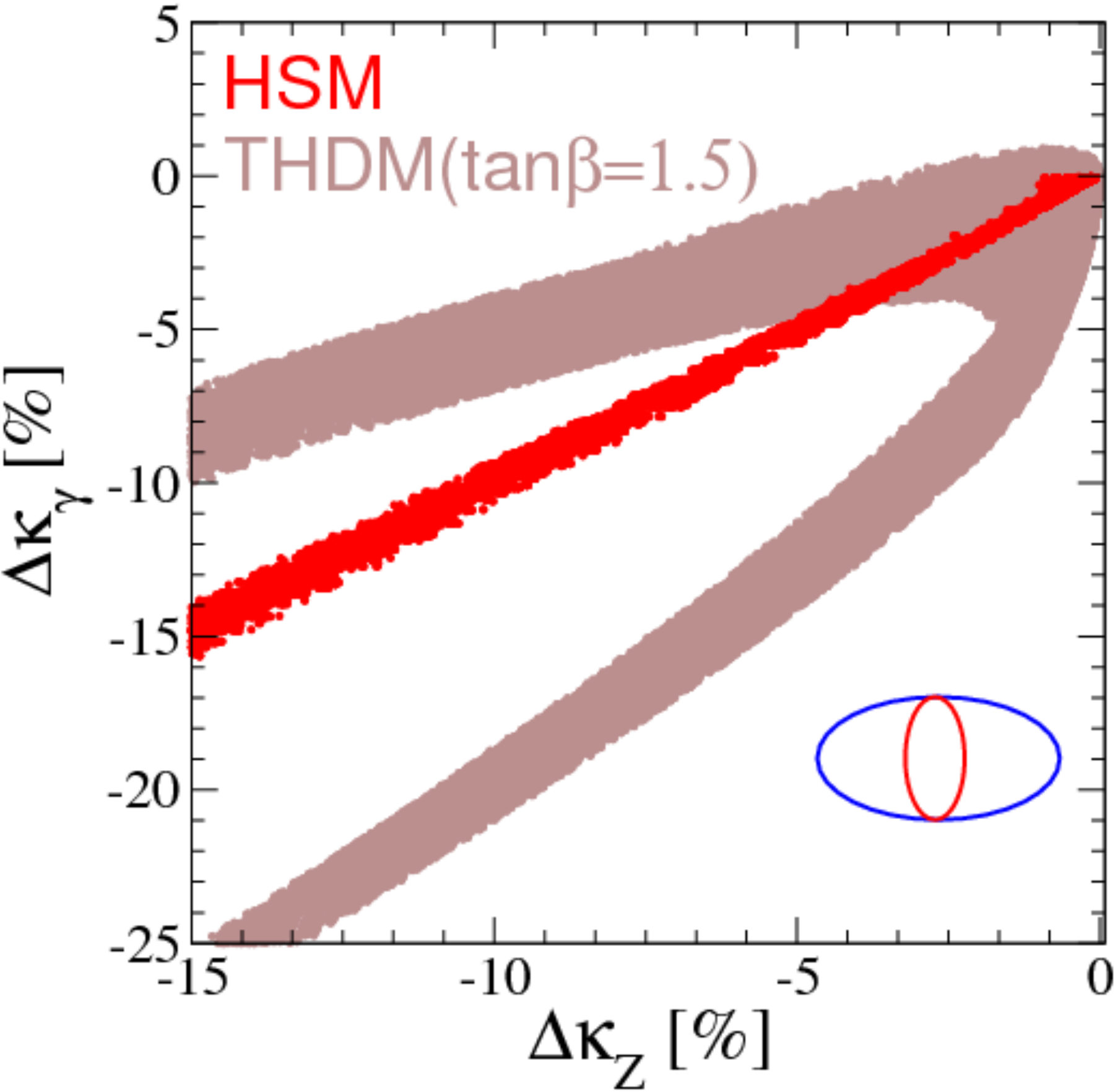} \vspace{5mm}\hspace{10mm}
  \includegraphics[width=6cm]{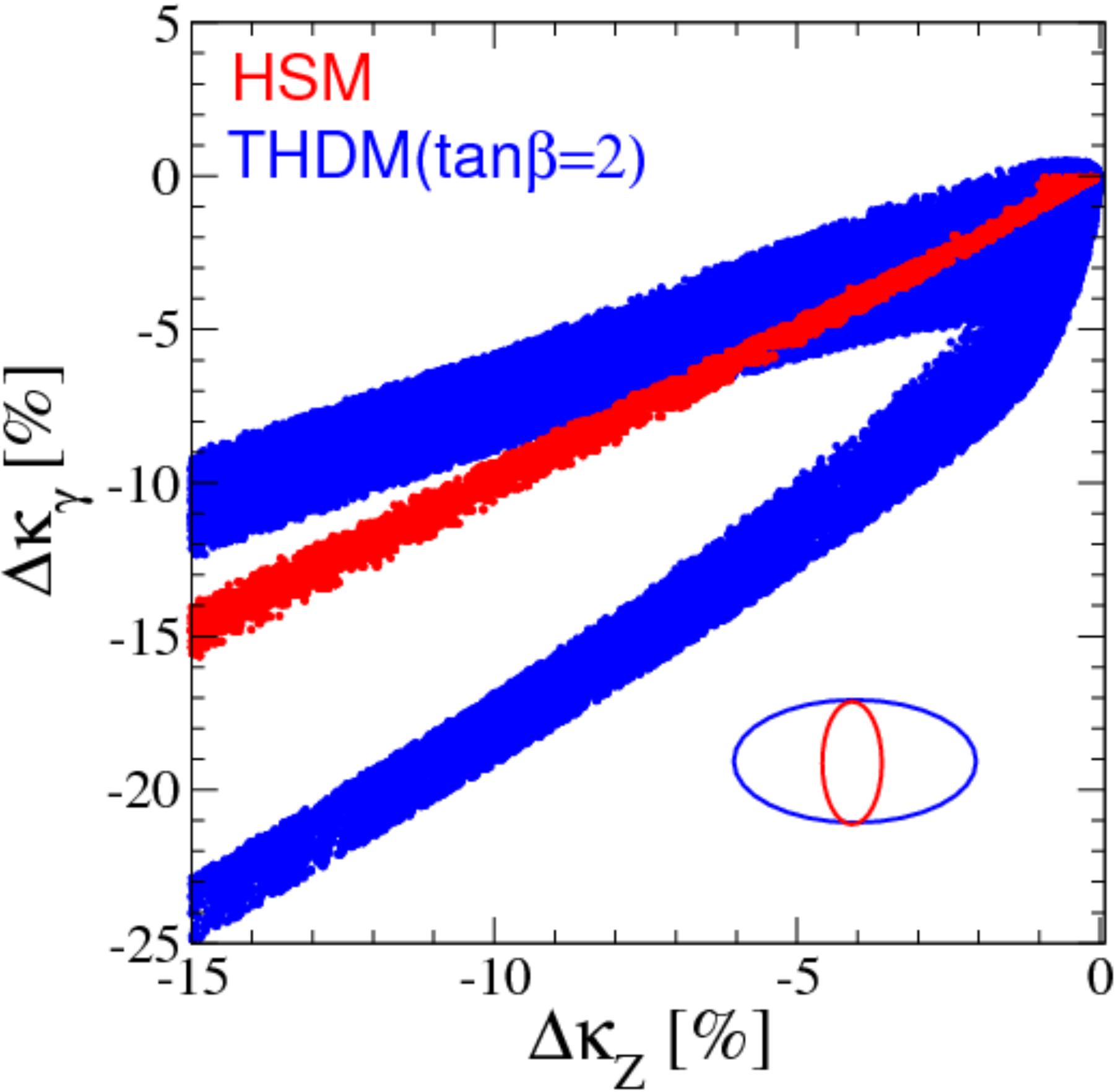} \vspace{5mm}\hspace{10mm}
  \includegraphics[width=6cm]{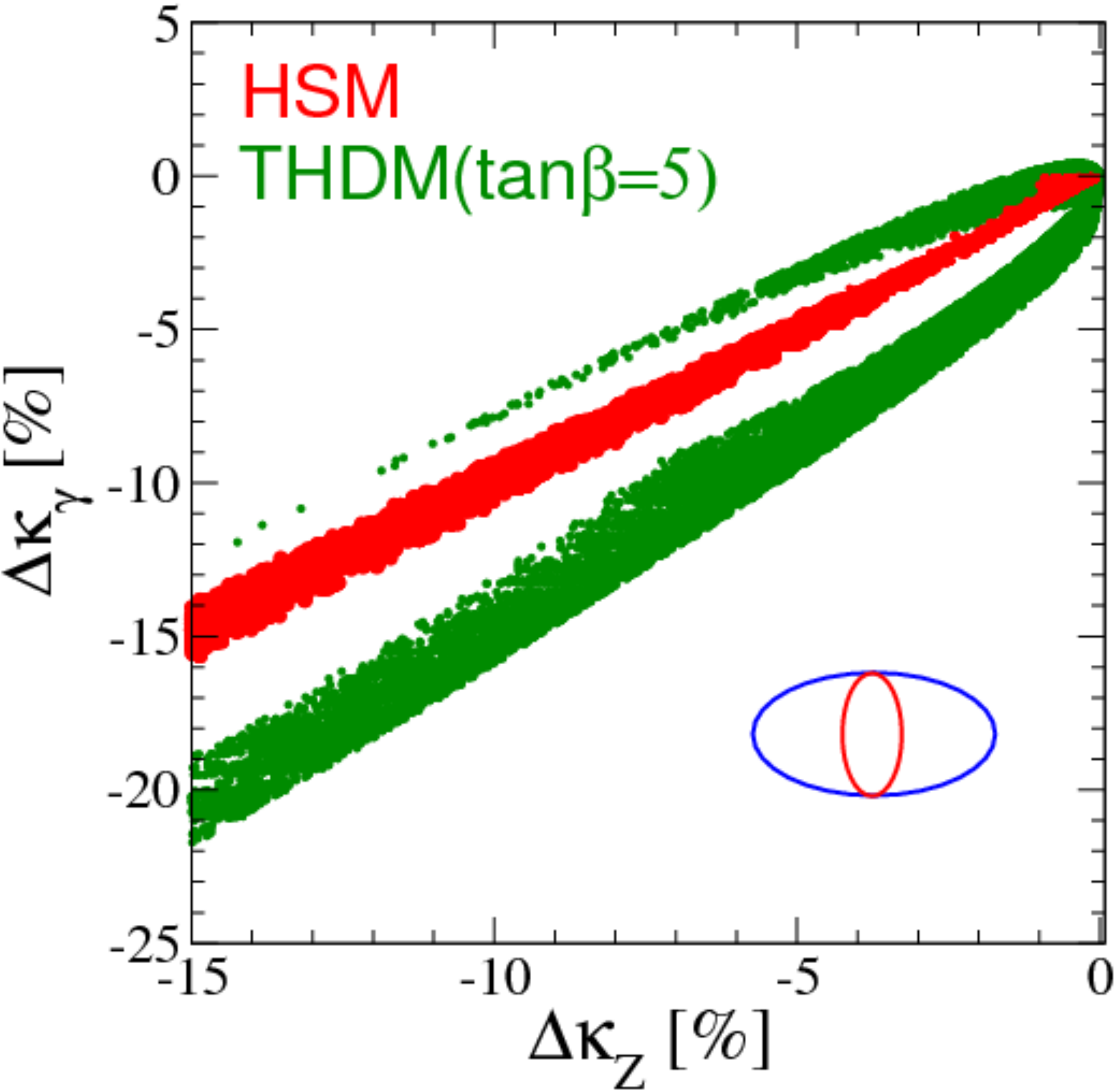} \vspace{5mm}\hspace{10mm}
  \includegraphics[width=6cm]{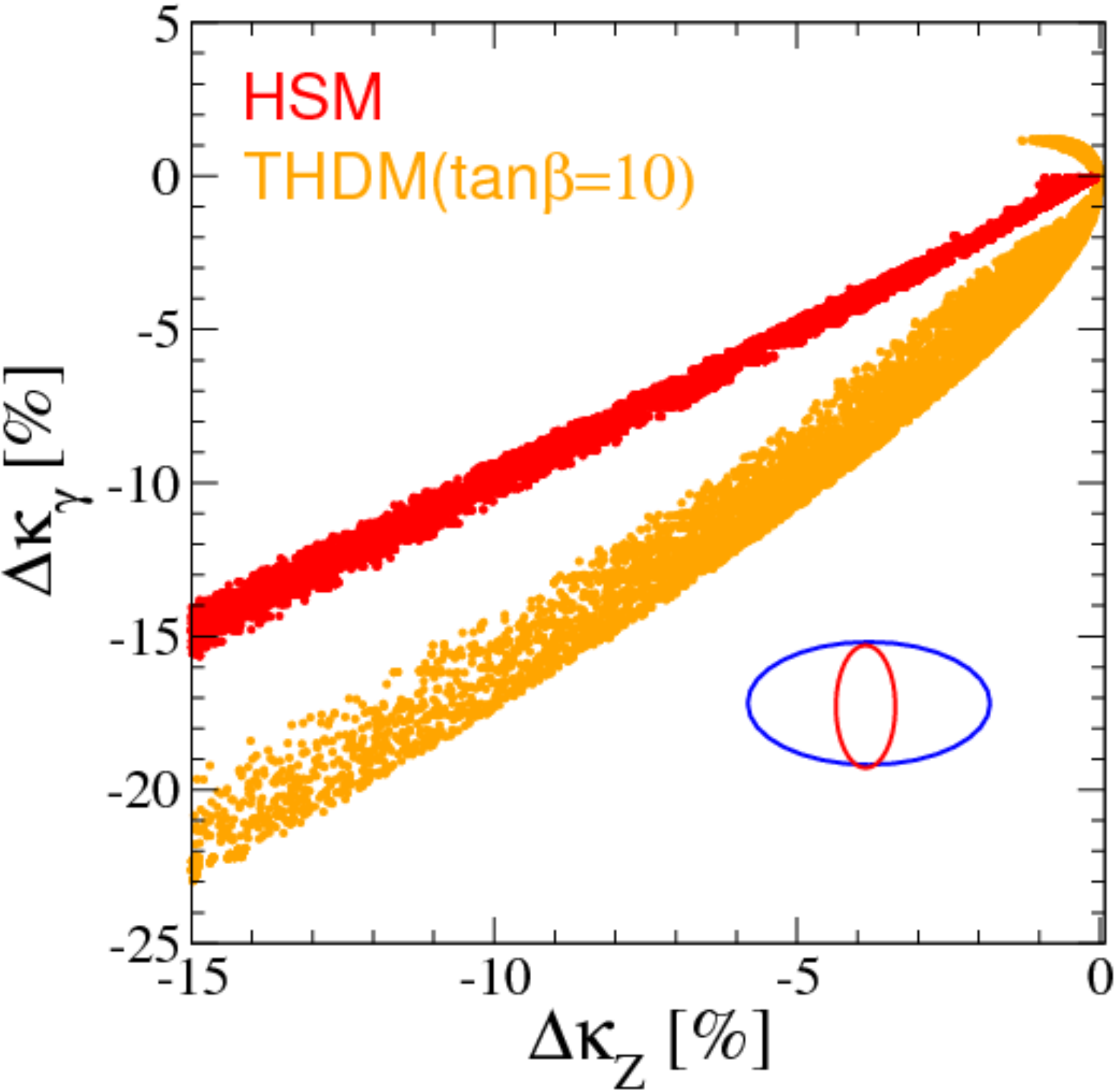} 
  \caption{Each panel shows the predictions of the allowed regions of the HSM and the Type I THDM on the $\Delta \kappa_Z^{}$ and $\Delta \kappa_\gamma^{}$ plane. The value of $\tan\beta$ is fixed to be 1.5 (upper left), 2 (upper right), 5 (bottom left), and 10 (bottom right).
The scanned range of the parameters are the same as that in Fig.~\ref{dkv_dkg}.}
  \label{dkv_dkg2}
 \end{figure}

In the HSM, the correlation between $\Delta \hat{\kappa}_Z^{}$ and $\Delta \kappa_\gamma^{}$ follows the line of $1:1$ with the small width, which comes from radiative corrections. 
Because there is no charged new particle in the HSM, deviations in $\Delta \kappa_\gamma^{}$ are made by mixing effects. 
In the THDM, in addition to mixing effects, singly charged Higgs bosons loop contributions modify the value of $\Delta \kappa_\gamma^{}$.
The magnitude of $\Delta \kappa_\gamma^{}$ depends on the sign of $\cos(\beta-\alpha)$ as the behavior of $\Delta\hat{\kappa}_b^{}$.
Predictions distributing in region of smaller $|\Delta \kappa_\gamma|$ values is the case for $\cos(\beta-\alpha)>0$.
The other regions show the predictions with $\cos(\beta-\alpha)<0$. 
In the limit for $\tan\beta \rightarrow \infty$, the predictions of the Type I THDM are close the line of $1:1$.
There are allowed regions with $\Delta\kappa_\gamma^{} > 0$, which are caused by inversing the sign of the $hH^+H^-$ coupling. 
If it is difficult to identify the results of each value of $\tan\beta$ in the plane of $\Delta \kappa_Z^{}$ and $\Delta \kappa_\gamma^{}$,
you can see their behavior more clearly by using Fig.~\ref{dkv_dkg2}.
In each panel of Fig.~\ref{dkv_dkg2}, we show allowed regions of $\Delta \hat{\kappa}_Z^{}$ and $\Delta \kappa_\gamma^{}$ in the HSM and the Type I THDM for each value of $\tan\beta$.
The definition of colors and ellipses, and the way of analysis are same as those in Fig.~\ref{dkv_dkb} and Fig.~\ref{dkv_dkg}.

Finally, we discuss how we can discriminate the HSM and the Type I THDM by using theoretical predictions of $\Delta \kappa_Z^{}$, $\Delta \kappa_b^{}$ and $\Delta \kappa_\gamma^{}$ with radiative corrections 
and Higgs boson coupling measurements at the HL-LHC and the ILC500.
We find that if $\kappa_V^{}$ will be measured to be deviated by 2 \% from the SM predictions, 
we can discriminate the HSM and the Type I THDM in most of parameter regions by using precision measurements of $\Delta \kappa_Z^{} $ and $\Delta \kappa_b^{}$ at the ILC.
In addition, in the plane of $\Delta \kappa_Z^{}$ and $\Delta \kappa_\gamma^{}$, the predictions of the HSM separate from those of the Type I THDM for $\cos(\beta-\alpha)>0$.
However, when the value of $\tan\beta$ is extremely large; i.e., $\tan\beta \gg 10$, 
$\Delta \kappa_Z^\textrm{THDM}$, $\Delta \kappa_b^\textrm{THDM}$ and $\Delta \kappa_\gamma^\textrm{THDM}$ approach to the predictions in the HSM. 
In such a situation, it is difficult to discriminate the models by only using these coupling constants.

\section{Discussion and Conlusion}
We have calculated a full set of renormalized Higgs boson couplings at the one-loop level in the on-shell scheme in the HSM.
These coupling constants can deviate from the SM predictions due to the mixing effect and the one-loop contributions of the extra scalar boson. 
We numerically have investigated how they can be significant under the theoretical constraints from perturbative unitarity and vacuum stability and also 
the condition of avoiding the wrong vacuum. 
Finally, comparing with the predictions at the one-loop level in the four types of THDMs, 
we have studied how the HSM can be distinguished from those models and identified by using precision measurements of the Higgs boson couplings 
at future collider experiments.
We found that if $hVV$ couplings deviate 2 \% from the SM predictions, 
we can discriminate the HSM and the Type I THDM in most of parameter regions by using precision measurements of $\Delta \kappa_Z^{} $ and $\Delta \kappa_b^{}$ at the ILC.
In addition that, in the plane of $\Delta \kappa_Z^{}$ and $\Delta \kappa_\gamma^{}$, the predictions of the HSM separate from those of the Type I THDM for $\cos(\beta-\alpha)>0$.
Therefore, by comparing the predicted values of the $hZZ$, $hb\bar{b}$ and $h\gamma\gamma$ couplings and corresponding measured values, 
we may be able to distinguish the HSM from the Type I THDM in the most of the parameter space. 
However, when the value of $\tan\beta$ is extremely large as $\tan\beta > 10$, 
deviations in $\Delta \kappa_Z^\textrm{THDM}$, $\Delta \kappa_b^\textrm{THDM}$ and $\Delta \kappa_\gamma^\textrm{THDM}$ approach to the predictions in the HSM. 
In such a situation, it is difficult to discriminate the models by fingerprinting.\\


\noindent
{\it Acknowledgments}

S.K. was supported in part by Grant-in-Aid for Scientific Research, The Ministry of Education, Culture, 
Sports, Science and Technology (MEXT), No. 23104006, and Grant H2020-MSCA-RISE-2014 No. 645722 (Non-Minimal Higgs). 
M.K. was supported in part by JSPS, No. K25$\cdot$10031.
K.Y. was supported by JSPS postdoctoral fellowships for research abroad.


\clearpage
 \appendix
\section{Theoretical constraints}\label{sec:constraints}
In this section, we summarize three theoretical constraints; i.e., perturbative unitarity, vacuum stability and the condition to avoid the wrong vacuum. 
\subsection{Perturbative unitarity}
The constraints from the perturbative unitarity in the HSM had discussed in Ref.~\cite{Uni}.
Under the perturbative unitarity bound, the matrix of the S-wave amplitude for the two-body to two-body scattering of scalar fields has to be satisfied in following conditions,
 \begin{align}
 |\langle\varphi_3\varphi_4|a_0|\varphi_1\varphi_2\rangle| < \xi,
 \,\,\,\, \textrm{where} \,\,\, \xi = 1 \,\,\, \textrm{or} \,\,\,
 \frac{1}{2}.
 \label{uni1}
 \end{align}
In the HSM, there are seven neutral scattering processes.\footnote{Although there are one doubly charged cannel and three singly charged cannels in addition to seven neutral channels, independent eigenvalues is exhausted in eigenvalues of neutral scattering amplitudes. Because of that, it is sufficient to consider only neutral cannels.}
Digonalizing the matrix of the neutral scattering processes, we obtain following independent eigenvalues,
  \begin{align}
 a_\pm &=  \frac{1}{16\pi}\left( 3\lambda + 6\lambda_S 
   \pm \sqrt{(3\lambda -6\lambda_S)^2 +4\lambda_{\Phi S}^2} \right) , \\
 b_0 &= \frac{1}{8\pi}\lambda, \\
 c_0 &= \frac{1}{8\pi}\lambda_{\Phi S}.
 \label{uni2}
 \end{align}
Because we take the constraint with $\xi=\frac{1}{2}$, specific bounds of Eq.~(\ref{uni1}) are
  \begin{align}
& \left( 3\lambda + 6\lambda_S 
   \pm \sqrt{(3\lambda -6\lambda_S)^2 +4\lambda_{\Phi S}^2} \right) <8\pi, \,\,\,\,\,
 \lambda < 4\pi, \,\,\,\,\, 
 \lambda_{\Phi S} < 4\pi.
 \end{align}

 \subsection{Vacuum stability}
As conditions of vacuum stability, 
we require the value of the potential to be positive at large $\Phi$ and $S$.
Because terms of the quartic interactions are dominant in the potential with large values of the fields,
 \begin{align}
 \lambda |\Phi|^4 + \lambda_{\Phi S}^{} |\Phi|^2 S^2 +\lambda_S^{} S^4 > 0
 \label{vacuum stability2}
 \end{align} 
must be satisfied.
In order to satisfy ~\ref{vacuum stability2}, following bounds for $\lambda$ parameters are imposed,
 \begin{align}
 \lambda >0 ,\,\,\,  \lambda_S >0, \,\,\, 
 4\lambda \lambda_S > \lambda_{\Phi S}^2,
 \end{align}
where the third bound is applied when $\lambda_{\Phi S}^{} $ is negative.

\subsection{To avoid the wrong vacuum}
We are free to choose the value of $v_S^{}$.
We take to be $(v, v_S^{}) = (v_{EW}, 0)$, because the singlet field does not contribute to electroweak symmetry breaking.
However, even $(v_{EW},0)$ is the extrema, there is a possibility that there are lower extremes at other points.
According to Refs~\cite{chen_dawson_lewis,wrong_vacuum}, five kinds of other extrema.
If one or more than one extrema given in Eq.~(24) and (B1) Ref.~\cite{chen_dawson_lewis} become deeper than $V(v_{EW}^{}, 0)$, then such a vacuum should be regarded as a wrong vacuum.
In the analyses of this paper, 
we use the condition to avoid the wrong vacuum given in Ref.~\cite{chen_dawson_lewis}.

\section{One-loop level corrected electroweak observables}\label{sec:STU}

We here list the renormalized electroweak parameter $\Delta r$ and renormalized $W$ boson mass $m_W^{\textrm{reno}}$.
They can be expressed as~\cite{EW_reno}
 \begin{align}
 \Delta r & = \frac{d}{dp^2}\Pi_{\gamma\gamma}^{1\PI}[p^2]\Bigg|_{p^2=0} 
  -\frac{c_W^2}{s_W^2}\left( 
   \frac{\RE \Pi_{ZZ}^{1\PI}[m_Z^2]}{m_Z^2} 
 - \frac{\RE \Pi_{WW}^{1\PI}[m_W^2]}{m_W^2}
 - \frac{2s_W^{}}{c_W^{}} \frac{\Pi_{\gamma Z}^{1\PI}[0]}{m_Z^2} \right) \notag\\ 
&+ \frac{\Pi_{WW}^{1\PI}[0] - \Pi_{WW}^{1\PI}[m_W^2]}{m_W^2}
 + \delta_{VB}, \\
 (m_W^{\textrm{reno}})^2 & = \frac{m_Z^2}{2}\left(
     1 + \sqrt{1 - \frac{4\pi \alpha_{\EM}^{}}{\sqrt{2}G_F^{}m_Z^2(1-\Delta r)}}
 \right), 
 \end{align}
where $\delta_{VB}$ is the box and the vertex diagram contributions to the muon decay process, which is given by~\cite{EW_reno} 
 \begin{align}
 \delta _{VB}^{} = \frac{\alpha _{\EM}^{}}{4\pi s_W^2}\left(
  6 + \frac{7 - 4s_W^2 }{2s_W^2}
  \ln \frac{m_W^2}{m_Z^2} \right).
 \end{align}
  
Moreover, we also can calculate electroweak $S$, $T$ and $U$ parameters as 
 \begin{align}
 S &= \frac{16\pi}{m_Z^2}\RE\Bigg[
      \frac{c_{2W}^{}}{eg_Z^{}}\biggl(
         \Pi_{Z\gamma}^{1\PI}[m_Z^2] - \Pi_{Z\gamma}^{1\PI}[0] 
            \bigg)
    +\frac{s_W^2c_W^2}{e^2} \left(
          \Pi_{\gamma\gamma}^{1\PI}[m_Z^2] - \Pi_{\gamma\gamma}^{1\PI}[0]  \right) 
  \notag\\
  &\,\,\,\,\,\,\,\,\,\,\,\,\,\,\,\,\,\,
    +\frac{1}{g_Z^2}\left(
          \Pi_{ZZ}^{1\PI}[0] - \Pi_{ZZ}^{1\PI}[m_Z^2]\right) \Bigg], \\
 T &= \frac{1}{\alpha_{\EM}}\RE\left[-\frac{\Pi_{WW}^{1\PI}[0]}{m_W^2} 
        +\frac{\Pi_{ZZ}^{1\PI}[0]}{m_Z^2}
        +2\frac{s_W^{}}{c_W^{}}\frac{\Pi_{\gamma Z}^{1\PI}[0]}{m_Z^2}
    +\frac{s_W^2}{c_W^2}\frac{\Pi_{\gamma \gamma}^{1\PI}[0]}{m_Z^2}\right], \\
 U  &=16\pi\RE\Biggl[
      -\frac{1}{m_Z^2}\Big\{\frac{1}{g_Z^2} \left(
          \Pi_{ZZ}^{1\PI}[0] - \Pi_{ZZ}^{1\PI}[m_Z^2]\ \right) 
       +\frac{2s_W^2}{eg_Z^{}}
           \left(\Pi_{Z\gamma}^{1\PI}[0] - \Pi_{Z\gamma}^{1\PI}[m_Z^2] \right)
  \notag\\
  &\,\,\,\,\,\,\,\,\,\,\,\,\,\,\,\,\,\,
            +\frac{s_W^4}{e^2}\left(
           \Pi_{\gamma\gamma}^{1\PI}[0] - \Pi_{\gamma\gamma}^{1\PI}[m_Z^2] \right) 
             +\frac{1}{g^2m_W^2}
       \left(\Pi_{WW}^{1\PI}[0] - \Pi_{WW}^{1\PI}[m_W^2] \right) \Big\}
    \Biggr], 
 \end{align}
where $g_Z^{} = g/c_W^{}$.

 \section{Tree level Higgs boson couplings}

First, we give feynman rules of trilinear vertices and quartic vertices obtained from the Higgs kinetic term.
There are two kinds of trilinear vertices and one kind of quartic vertices; i.e., Scalar-Gauge-Gauge, Scalar-Scalar-Gauge and Scalar-Scalar-Gauge-Gauge type.
Their couplings are expressed as
 \begin{align}
 \mathcal{L} =   g_{\phi V_1 V_2}^{}g^{\mu\nu}\phi V_{1\mu}^{}V_{2\nu}^{}
               + g_{\phi_1\phi_2  V}^{}
                (\partial^\mu\phi_1^{}\phi_2^{} - \phi_1^{}\partial^\mu\phi_2^{})V_{\mu}^{}
               + g_{\phi_1\phi_2 V_1 V_2}^{}g^{\mu\nu}\phi_1\phi_2 V_{1\mu}^{}V_{2\nu}^{} + \cdots.
 \end{align} 
The coefficients of trilinear vertices $g_{\phi V_1 V_2}^{}$ and $g_{\phi_1 \phi_2 V}^{\mu}$, and those of quartic vertices $g_{\phi_1\phi_2V_1V_2}^{}$ are listed in Tab.~\ref{s_g_3} and in Tab.~\ref{s_g_4}, where $p_1^\mu$ ($p_2^{\mu}$) indicates incoming momentum of $\phi_1$ ($\phi_2$).   

  \begin{table}[htb]
  \renewcommand{\arraystretch}{1.5}
   \centering
   \caption{The Scalar-Vector-Vector vertices and the Scalar-Scalar-Vector vertices and those coefficients. }
   \begin{tabular}{cc||cc}
   \hline
    $\phi V_{1\mu}^{} V_{2\nu}^{} $ vertices & coefficient &
    $\phi_1^{} \phi_2^{} V_\mu $ vertices  &  coefficient \\\hline
    $hW^+_\mu W^-_\nu$ & $\frac{2m_W^2}{v}c_\alpha^{}$ &
    $hG^\pm W^\mp_\mu$ & $\mp i\frac{m_W}{v}c_\alpha$ \\
    $HW^+_\mu W^-_\nu$ & $\frac{2m_W^2}{v}s_\alpha$ &
    $HG^\pm W^\mp_\mu$ & $\mp i\frac{m_W}{v}s_\alpha$ \\
    $G^\pm Z_\mu W^\mp_\nu$ & $-\frac{2m_W m_Z}{v}s_W^2 $ &
    $G^0G^\pm W^\mp_\mu$ & $-\frac{m_W}{v}$ \\
    $G^\pm \gamma_\mu W^\mp_\nu$ & $e m_W$ &
    $G^+ G^- Z_\mu^{}$ & $ i\frac{m_Z}{v}c_{2W}$ \\
    $hZ_\mu^{} Z_\nu^{}$ & $\frac{m_Z^2}{v}c_\alpha$ &
    $h G^0 Z_\mu^{}$ & $ -\frac{m_Z}{v}c_\alpha$ \\
    $HZ_\mu^{}Z_\nu^{}$ & $\frac{m_Z^2}{v}s_\alpha$ &
    $HG^0 Z_\mu^{}$ & $ -\frac{m_Z}{v}s_\alpha  $ \\
    & &
    $G^+ G^- \gamma_\mu^{}$ & $ ie$ \\
   \hline
   \end{tabular}
   \label{s_g_3}
  \end{table}

  \begin{table}[htb]
  \renewcommand{\arraystretch}{1.5}
   \centering
   \caption{The Scalar-Scalar-Vector-Vector vertices and those coefficients. }
   \vspace{2mm}
   \begin{tabular}{cc||cc}
   \hline
    $\phi_1^{} \phi_2^{} V_{1\mu} V_{2\nu} $ vertices  & coefficient &
    $\phi_1^{} \phi_2^{} V_{1\mu} V_{2\nu} $ vertices  & coefficient \\\hline
    $hhW^+_\mu W_\nu^-$ & $\frac{m_W^2}{v^2}c_\alpha^2$ &
    $G^\pm G^0W^\mp_\mu Z_\nu^{}$ & $\pm i\frac{2m_W m_Z}{v^2}s_W^2$ \\
    $HHW^+_\mu W_\nu^-$ & $\frac{m_W^2}{v^2}s_\alpha^2$ &
    $G^\pm h W^\mp_\mu Z_\nu^{}$ & $-\frac{2m_W m_Z}{v^2}s_W^2c_\alpha $ \\
    $G^0 G^0W^+_\mu W_\nu^-$ & $\frac{m_W^2}{v^2}$ &
    $G^\pm H W^\mp_\mu Z_\nu^{}$ & $-\frac{2m_W m_Z}{v^2}s_W^2s_\alpha $ \\
    $G^+ G^+W^+_\mu W_\nu^-$ & $\frac{2m_W^2}{v^2}$ &
    $G^\pm h W^\mp_\mu \gamma_\nu^{}$ & $\frac{em_W}{v}c_\alpha$ \\
    $hhZ_\mu^{}Z_\nu^{}$ & $\frac{m_Z^2}{2v^2}c_\alpha^2$ &
    $G^\pm H W^\mp_\mu \gamma_\nu^{}$ & $\frac{em_W}{v}s_\alpha$ \\
    $HHZ_\mu^{}Z_\nu^{}$ & $\frac{m_Z^2}{2v^2}s_\alpha^2$ &
    $G^\pm G^0 W^\mp_\mu \gamma_\nu^{}$ & $\mp i\frac{em_W}{v}$ \\
    $G^0 G^0Z_\mu^{}Z_\nu^{}$ & $\frac{m_Z^2}{2v^2}$ &
    $G^+ G^- Z_\mu^{} \gamma_\nu^{}$ & $ 2e\frac{m_Z}{v}c_{2W}$ \\
    $G^+ G^- Z_\mu^{} Z_\nu^{}$ & $\frac{m_Z^2}{v^2}c_{2W}^2$ &
    $G^+ G^- \gamma_\mu^{} \gamma_\mu^{}$ & $ e^2$ \\
    $hHW^+_\mu W^-_\mu$ & $\frac{2m_W^2}{v^2}s_\alpha c_\alpha$ &
    $hHZ_\mu^{}Z_\mu^{} $ & $ \frac{m_Z^2}{v^2}s_\alpha c_\alpha$ \\
   \hline
   \end{tabular}
  \label{s_g_4}
  \end{table}

We give feynman rules of the scalar trilinear and the quartic vertices.
When we express those couplings as
 \begin{align}
 \mathcal{L} =
                \lambda_{\phi_1^{} \phi_2^{} \phi_3^{}}^{} 
                \phi_1^{} \phi_2^{} \phi_3^{}
              + \lambda_{\phi_1^{} \phi_2^{} \phi_3^{} \phi_4^{}}^{} 
                \phi_1^{} \phi_2^{} \phi_3^{} \phi_4^{}
              + \cdots,
 \end{align}
those coefficients $\lambda_{\phi_1^{} \phi_2^{} \phi_3^{}}^{}$ and $\lambda_{\phi_1^{} \phi_2^{} \phi_3^{} \phi_4^{}}^{}$ are obtained as
 \begin{align}
 \renewcommand{\arraystretch}{1.5}
 \lambda_{hhh} & = -\frac{c_\alpha^3}{2v}m_h^2
                 - s_\alpha^2( c_\alpha^{}\lambda_{\Phi S}v
                            -4s_\alpha^{} \lambda_S^{}v_S^{}
                            - s_\alpha^{} \mu_S^{} ), \\
  \lambda_{hHH} &= 
 -(m_h^2 + 2m_H^2)\frac{c_\alpha^{}s_\alpha^{2}}{2v}
 - \frac{\lambda_{\Phi S} v}{4} (c_\alpha^{} + 3c_{3\alpha}^{} )
 + 12s_\alpha^{} c_{\alpha}^2 \lambda_S^{} v_S^{}
 + 3s_\alpha^{} c_\alpha^2 \mu_S^{}, \\
 \lambda_{hhH} &= 
 - (2m_h^2 + m_H^2 )\frac{s_\alpha c_\alpha^2}{2v}
 + \frac{s_\alpha \lambda_{\Phi S}^{} v}{2}(1 + 3c_{2\alpha}^{})
 - 3s_\alpha^2 c_\alpha^{} \mu_S^{}
 - 6s_\alpha^{} s_{2\alpha}^{}\lambda_S^{} v_S^{},\\ 
 \lambda_{HHH} &=
 - \frac{s_\alpha^3}{2v}m_H^2 
 - 4c_\alpha^3 \lambda_S^{} v_S^{} - c_\alpha^3 \mu_S^{}
 - s_\alpha^{} c_\alpha^{2} \lambda_{\Phi S}^{} v, \\
 \lambda_{hG^0 G^0} &= -\frac{m_h^2 c_\alpha}{2 v}, \\
 \lambda_{hG^+ G^-}&= -\frac{m_h^2 c_\alpha}{ v},  \\
 \lambda_{HG^0 G^0} &= -\frac{m_H^2 s_\alpha}{2 v}, \\
 \lambda_{HG^+ G^-} &=-\frac{m_H^2 s_\alpha}{v}, 
 \end{align}
 \begin{align}
 \lambda_{hhhh}^{} &=
 - (c_\alpha^2 m_h^2 + s_\alpha^2 m_H^2) \frac{c_\alpha^4}{8v^2}
 - s_\alpha^4 \lambda_S^{}
 - \frac{s_{2\alpha}^2}{8}\lambda_{\Phi S}^{}, \\
 \lambda_{hhhH}^{} & =
 - \frac{c_\alpha^5 s_\alpha^{}}{2v^2} m_h^2 
 - \frac{s_{2\alpha}^3}{16v^2}m_H^2
 + 4 c_\alpha^{} s_{\alpha}^3 \lambda_S^{}
 + \frac{s_{4\alpha}^{}}{4} \lambda_{\Phi S}^{}, \\
 \lambda_{hhHH}^{} & =
 - (c_\alpha^2 m_h^2 + s_\alpha^{2} m_H^2)\frac{3s_\alpha^2 c_\alpha^2}{4v^2}
 - \frac{\lambda_{\Phi S}^{}}{8} (1+ 3c_{4\alpha}^{}), \\
 \lambda_{hHHH}^{} & =
 4\lambda_S^{}c_\alpha^3 s_\alpha^{}
 - \frac{m_H^2}{2v^2}c_\alpha s_\alpha^5
 - \frac{m_h^2}{16v^2}s_{2\alpha}^3 - \frac{\lambda_{\Phi S}^{}}{4}s_{4\alpha}^{},\\
 \lambda_{hhG^+G^-} &=
 - \frac{c_\alpha^4}{2v^2} m_h^2 - \frac{s_{2\alpha}^2}{8v^2} m_H^2
 - s_\alpha^2 \lambda_{\Phi S}^{}, \\
 \lambda_{hHG^+ G^-} & =
 - (c_\alpha^2 m_h^2 + s_\alpha^2 m_H^2) \frac{s_\alpha^{}c_\alpha^{} }{v^2}
 + 2 s_\alpha^{} c_\alpha^{} \lambda_{\Phi S}^{}, \\
 \lambda_{HHG^+ G^-} &=
 - (4s_\alpha^4 m_H^2 + m_h^2 s_{2\alpha}^2) \frac{1}{8v^2}
 - c_\alpha^2 \lambda_{\Phi S}^{}, \\
 \lambda_{hhG^0G^0}^{} &=
 - \frac{m_h^2}{2v^2}c_\alpha^4 - \frac{m_H^2}{16v^2}s_{2\alpha}^2
 - \frac{\lambda_{\Phi S}}{2} s_\alpha^2, \\
 \lambda_{hHG^0G^0} &=
 - \frac{m_h^2}{2v^2}c_\alpha^3 s_\alpha^{}
 - \frac{m_H^2}{2v^2}c_\alpha^{} s_\alpha^{3} + \lambda_{\Phi S}c_\alpha^{}s_\alpha^{}.
 \end{align}

 \section{1PI diagrams}\label{sec:1PI}

In this section, we give one-loop fermion, vector boson and scalar boson contributions to the one, two and three point functions by using Passarino-Veltman functions~\cite{PV_func} whose notation is same as those defined in Ref.~\cite{Hagiwara}.
We calculate 1PI diagrams in the 't~Hooft--Feynman gauge so that the masses of Numbu-Goldstone bosons $m_{G^\pm}$ and $m_{G^0}$ and those of Fadeev-Popov ghosts $m_{c^{\pm}}$, $m_{c^0}$ and $m_{c^\gamma}^{}$ are the same as corresponding masses of the gauge bosons.
We write 1PI diagram contributions separately for fermion loop contributions and boson loop contributions which are expressed by index $F$ and $B$, respectively.

 \subsection{One point functions}
The 1PI tadpole contributions are calculated by
 \begin{align}
  (16\pi^2)  T_h^{1\PI ,F}& = 
    -\sum _{f} 4 \frac{m_f^2}{v} c_\alpha N_c^f A(m_f), \\
  (16\pi^2)  T_H^{1\PI,F} &=
    -\sum _{f} 4 \frac{m_f^2}{v} s_\alpha N_c^f A(m_f), \\
  (16\pi^2) T_h^{1\PI ,B} &= 
    - 3\lambda_{hhh} A(m_h) 
    -\lambda_{hHH} A(m_H)
    -\lambda_{hG^+G^-} A(m_{G^\pm})
    -\lambda_{hG^0G^0} A(m_{G^0}) \notag\\
  & -g m_W c_{\alpha} A(m_{c^\pm})
    -\frac{g_Z m_Z}{2}c_{\alpha} A(m_{c^0}) 
    +\frac{2m_W^2}{v}c_{\alpha}D A(m_W)
    +\frac{m_Z^2}{v}c_{\alpha} D A(m_Z) 
   , 
 \end{align}
 \begin{align}
  (16\pi^2)  T_H^{1\PI,B} &=
    -\lambda_{hhH} A(m_{h}) 
    -3\lambda_{HHH} A(m_H) 
    -\lambda_{HG^+G^-} A(m_{G^+})
    -\lambda_{Hzz} A(m_{G^0}) \notag\\
  & -g m_W s_{ \alpha} A(m_{c^\pm})
    -\frac{g_Z m_Z}{2} s_{\alpha} A(m_{c^0}) 
    +\frac{2m_W^2}{v} s_{\alpha} D A(m_W)
    +\frac{m_Z^2}{v} s_{\alpha} D A(m_Z), 
 \end{align}
where $D= 4- 2\epsilon$ and $N_c^f$ indicates the color number of each particle.

 \subsection{Two point functions}\label{sec:1PI 2point}

The 1PI diagram contributions to the scalar boson two point functions are expressed as 
 \begin{align}
  (16\pi^2)  \Pi_{hh}^{1\PI,F}[p^2] &=
     -2\sum_f 
     \left( \frac{m_f}{v} c_\alpha \right)^2N_c^f
    \left\{ 2A(m_f) -(p^2-4m_f^2)B_0(p^2; m_f, m_f) \right\},\\
  (16\pi^2)  \Pi_{Hh}^{1\PI,F}[p^2] &=
    -2\sum_f 
    \left( \frac{m_f}{v}  \right)^2 c_\alpha s_\alpha N_c^f
    \left\{ 2A(m_f) -(p^2-4m_f^2)B_0(p^2; m_f, m_f) \right\},\\
  (16\pi^2)  \Pi_{HH}^{1\PI,F}[p^2] &=
     -2\sum_f 
     \left( \frac{m_f}{v} s_\alpha \right)^2N_c^f
    \left\{ 2A(m_f) -(p^2-4m_f^2)B_0(p^2; m_f, m_f) \right\}, \\
  (16\pi^2)  \Pi_{hh}^{1\PI,B}[p^2] &=
    -2\lambda_{hhHH} A(m_H)
    -12\lambda_{hhhh} A(m_h)
    -2\lambda_{hhG^+G^-} A(m_{G^\pm}) \notag\\
  & -2\lambda_{hhG^0G^0} A(m_{G^0})
    +2 \frac{m_W^2}{v^2} c_\alpha^2 D A(m_W)
    + \frac{m_Z^2}{v^2} c_\alpha^2 D A(m_Z) \notag\\
  & + \lambda_{hG^+G^-}^2 B_0(p^2; m_{G^+}, m_{G^-})
    + 2\lambda_{hHH}^2 B_0(p^2; m_H, m_H) \notag\\
  & +18\lambda_{hhh}^2 B_0(p^2; m_h, m_h) 
    +4 \lambda_{hhH}^2 B_0(p^2; m_h, m_H) \notag\\
  & +2 \lambda_{hG^0 G^0}^2 B_0(p^2; m_{G^0}, m_{G^0}) 
    +\frac{4m_W^4}{v^2} c_{\alpha}^2 D B_0(p^2; m_W, m_W) \notag\\
  & +\frac{2m_Z^4}{v^2} c_{\alpha}^2 D B_0(p^2; m_Z, m_Z) \notag\\
  & -\frac{2m_W^2}{v^2} c_{\alpha}^2 \left\{
      2A(m_W) - A(m_{G^\pm}) + (2p^2-m_W^2+2m_{G^\pm}^2) B_0(p^2; m_W, m_{G^\pm}) 
                      \right\} \notag\\
  & -\frac{m_Z^2}{v^2} c_{\alpha}^2 \left\{
      2A(m_Z) - A(m_{G^0}) + (2p^2-m_Z^2+2m_{G^0}^2) B_0(p^2; m_Z, m_{G^0}) 
                      \right\} \notag\\
  & -\frac{2m_W^4}{v^2}c_{\alpha}^2 B_0(p^2; m_{c^\pm}, m_{c^\pm})
    -\frac{m_Z^4}{v^2}c_{\alpha}^2 B_0(p^2; m_{c^0}, m_{c^0}),
 \end{align}
 \begin{align}
 (16\pi^2) \Pi_{Hh}^{1\PI,B}[p^2] &=
    - \lambda_{hHG^+G^-} A(m_{G^\pm})
    - 3 \lambda_{hHHH} A(m_H) 
    - 3 \lambda_{hhhH} A(m_h)
    - \lambda_{hHG^0G^0} A(m_{G^0}^{}) \notag\\
  & + 4D\frac{m_W^2}{v^2}s_\alpha c_\alpha A(m_W^2)
    + 2D\frac{m_Z^2}{v^2}s_\alpha c_\alpha A(m_Z^2) \notag\\
  & + \lambda_{hG^+G^-} \lambda_{HG^+G^-} B_0(p^2; m_{G^\pm}, m_{G^\pm})
    +6 \lambda_{hHH} \lambda_{HHH} B_0(p^2; m_H, m_H) \notag\\
  & +4 \lambda_{hhH} \lambda_{hHH} B_0(p^2; m_h, m_H)
    +6 \lambda_{hhh} \lambda_{hhH} B_0(p^2; m_h, m_h) \notag\\
  & +2 \lambda_{hG^0G^0} \lambda_{HG^0G^0} B_0(p^2; m_{G^0}, m_{G^0}) \notag\\
  & + \frac{4m_W^4}{v^2} s_{\alpha} c_{\alpha} D B_0(p^2; m_W, m_W)
    + \frac{2m_Z^4}{v^2} s_{\alpha} c_{\alpha} D B_0(p^2; m_Z, m_Z)
    \notag\\
  & - \frac{2m_W^2}{v^2}s_{\alpha} c_{\alpha} 
     \left\{ 2A(m_W) -A(m_{G^\pm})+ (2p^2 - m_W^2 + 2m_{G^\pm}^2)
          B_0(p^2; m_W, m_{G^\pm}))\right\} \notag\\
  & - \frac{m_Z^2}{v^2}s_{\alpha} c_{\alpha} 
     \left\{ 2A(m_Z) -A(m_{G^0})+ (2p^2 - m_Z^2 + 2m_{G^0}^2)
          B_0(p^2; m_Z, m_{G^0}))\right\} \notag\\
  & - 2\frac{m_W^4}{v^2}s_{\alpha} c_{\alpha}
          B_0(p^2; m_{c^\pm}, m_{c^\pm})
    - \frac{m_Z^4}{v^2}s_{\alpha} c_{\alpha}
          B_0(p^2; m_{c^0}, m_{c^0}), \\
  (16\pi^2)  \Pi_{HH}^{1\PI,B}[p^2] &=
    -12\lambda_{HHHH} A(m_H)
    -2\lambda_{hhHH} A(m_h)
    -2\lambda_{HHG^+G^-} A(m_{G^\pm}) \notag\\
  & -2\lambda_{HHG^0G^0} A(m_{G^0})
    +2 \frac{m_W^2}{v^2} s_\alpha^2 D A(m_W)
    + \frac{m_Z^2}{v^2} s_\alpha^2 D A(m_Z) \notag\\
  & + \lambda_{HG^+G^-}^2 B_0(p^2; m_{G^\pm}, m_{G^\pm})
    + 18\lambda_{HHH}^2 B_0(p^2; m_H, m_H) \notag\\
  & +2\lambda_{hhH}^2 B_0(p^2; m_h, m_h) 
    +4 \lambda_{hHH}^2 B_0(p^2; m_h, m_H) \notag\\
  & +2 \lambda_{HG^0 G^0}^2 B_0(p^2; m_{G^0}, m_{G^0}) 
    +\frac{4m_W^4}{v^2} s_{\alpha}^2 D B_0(p^2; m_W, m_W) \notag\\
  & +\frac{2m_Z^4}{v^2} s_{\alpha}^2 D B_0(p^2; m_Z, m_Z) \notag\\
  & -\frac{2m_W^2}{v^2} s_{\alpha}^2 \left\{
      2A(m_W) - A(m_{G^\pm}) + (2p^2-m_W^2+2m_{G^\pm}^2) B_0(p^2; m_W, m_{G^\pm}) 
                      \right\} \notag\\
  & -\frac{m_Z^2}{v^2} s_{\alpha}^2 \left\{
      2A(m_Z) - A(m_{G^0}) + (2p^2-m_Z^2+2m_{G^0}^2) B_0(p^2; m_Z, m_{G^0}) 
                      \right\} \notag\\
  & -\frac{2m_W^4}{v^2}s_{\alpha}^2 B_0(p^2; m_{c^\pm}, m_{c^\pm})
    -\frac{m_Z^4}{v^2}s_{\alpha}^2 B_0(p^2; m_{c^0}, m_{c^0}). 
 \end{align}

The fermion loop contributions to the gauge boson two point functions are calculated as 
 \begin{align}
  (16\pi^2)  \Pi_{WW}^{1\PI,F}[p^2] &=  
       \sum_f \frac{4m_W^2 N_c^f}{v^2} 
           \left[-B_4 + 2p^2 B_3\right](p^2; m_f, m_{f'}),\\
  (16\pi^2)  \Pi_{ZZ}^{1\PI,F}[p^2] &= 
       \sum_f
       \frac{4m_Z^2 N_c^f}{v^2} \left[
          2p^2(4s_W^4 Q_f^2 -4s_W^2 Q_f I_f + 2I_f^2) B_3
         -2I_f^2 m_f^2 B_0 
                              \right](p^2; m_f, m_f),\\
  (16\pi^2)  \Pi_{\gamma Z}^{1\PI,F}[p^2] &= 
      -\sum_f  
      \frac{4e m_Z N_c^f}{v}p^2 (-4s_W^2 Q_f^2 +2I_f Q_f)B_3(p^2; m_f, m_f),\\
  (16\pi^2)  \Pi_{\gamma\gamma}^{1\PI,F}[p^2] &=
      \sum_f
       8 e^2 N_c^f Q_f^2 p^2 B_3(p^2; m_f, m_f), 
 \end{align}
where $B_3(p^2; m_1, m_2) = -B_1(p^2; m_1, m_2) - B_{21}(p^2;m_1,m_2)$ and 
$B_4(p^2; m_1, m_2) = -m_1^2 B_1(p^2;m_2,m_1) - m_2^2B_2(p^2; m_1,m_2)$ defined in Ref.~\cite{Hagiwara} and $Q_f^{}$ is the electric charge of a fermion $f$.
The boson loop contributions to the gauge boson two point functions are calculated as 
  \begin{align}
  (16\pi^2) \Pi_{WW}^{1\PI,B}[p^2] &=
      \frac{m_W^2}{v^2} \Biggl[  
                s_{\alpha}^2B_5(p^2,m_{G^\pm},m_H) 
               +4s_{\alpha}^2 m_W^2 B_0(p^2,m_{W},m_H) \notag\\
         &  +4c_{\alpha}^2 m_W^2 B_0(p^2; m_W, m_h)
           +c_{\alpha}^2 B_5(p^2;m_{G^\pm},m_h) \notag\\
          & -4\left\{ (8c_W^2p^2 -(1-4s_W^2)m_W^2 -m_Z^2)B_0
                      -(\frac{9}{4}-2s_W^2)B_5\right\}(p^2; m_W, m_Z) \notag\\
          & -4\left\{ 2s_W^2\Big[(4p^2-2m_W^2)B_0-B_5\Big](p^2; m_W, m_\gamma)
                    +\frac{2p^2}{3} \right\}
                        \Biggr], \\
  (16\pi^2) \Pi_{ZZ}^{1\PI,B}[p^2] &= 
      \frac{m_Z^2}{v^2}\Biggl[
          s_{\alpha}^2 B_5(p^2,m_H,m_{G^0})
      + 4 s_{\alpha}^2 m_Z^2 B_0(p^2;m_Z,m_H) \notag\\
 & +4c_{\alpha}^2 m_Z^2 B_0(p^2; m_Z, m_h)
   + c_{\alpha}^2 B_5 (p^2; m_{G^0}, m_h) \notag\\ 
 & -4\left[ \left(\frac{23}{4}p^2 - 2m_W^2 \right)B_0 + 9p^2 B_3\right](p^2; m_W, m_W) \notag\\
  &  -4\frac{2p^2}{3} 
    +8s_W^2 p^2 \left[ \frac{11}{2}B_0 + 10B_3 \right](p^2; m_W, m_W) 
    +\frac{16s_W^2 p^2}{3}   \notag\\
  & - 4 s_W^4 p^2 \left[5B_0 + 12 B_3 \right](p^2; m_W, m_W)
    -\frac{8s_W^4 p^2}{3}   
                       \Biggr], \\
  (16\pi^2)  \Pi_{\gamma Z}^{1\PI,B}[p^2] &=  
   -\frac{2em_Z}{v} p^2 \Bigg\{
     \left[\frac{11}{2}B_0 + 10 B_3\right](p^2; m_W, m_W) \notag\\  
  & -s_W^2 \left[5B_0 + 12B_3\right](p^2; m_W, m_W) 
  +\frac{2}{3}(1-s_W^2)
   \Bigg\},  \\
  (16\pi^2)  \Pi_{\gamma\gamma}^{1\PI,B}[p^2] &= 
   -e^2p^2\bigg\{ (5B_0+12B_3)(p^2;m_W, m_W) +\frac{2}{3} \bigg\},
 \end{align}
where $B_5(p^2; m_1, m_2) = A(m_1) + A(m_2) - 4B_{22}^{}(p^2; m_1, m_2)$~\cite{Hagiwara}.

Next, we give one-loop contributions to fermion two point functions, which are composed of following three kind parts,
 \begin{align}
 \Pi_{ff}^{1\PI}[p^2] = m_f \Pi_{ff,S}^{1\PI}[p^2] + \slashed{p} \Pi_{ff,V}^{1\PI}[p^2] -
                     \slashed{p}\gamma_5 \Pi_{ff,A}^{1\PI}[p^2].
 \end{align} 
They are calculated as
 \begin{align}
 (16\pi^2)\Pi_{ff,S}^{1\PI}[p^2]&=
   -2\frac{m_Z^2}{v^2}(v_f^2-a_f^2)(2B_0[p^2;m_f,m_Z] - 1)
   -2(Q_f e)^2 (2B_0[p^2;m_f,m_\gamma] -1) \notag\\[+5pt]
 & +\frac{m_f^2}{v^2}c_\alpha^2 B_0[p^2;m_f,m_h] 
   +\frac{m_f^2}{v^2}s_\alpha^2 B_0[p^2;m_f,m_H] \notag\\[+5pt]
 & -\frac{m_f^2}{v^2} B_0[p^2;m_f,m_{G^0}] 
   -2 \frac{m_{f'}^2}{v^2} B_0[p^2;m_f,m_{G^\pm}],
  \\[+10pt]
 (16\pi^2)\Pi_{ff,V}^{1\PI}[p^2]&=
   -\frac{m_W^2}{v^2}(2B_1[p^2;m_{f'},m_W]+1)
   -\frac{m_Z^2}{v^2}(v_f^2+a_f^2)(2B_1[p^2;m_f,m_Z]+1) \notag\\[+5pt]
 & -(Q_f e)^2 (2B_1[p^2;m_f,m_\gamma]  + 1)
   -\frac{m_f^2}{v^2}c_\alpha^2 B_1[p^2;m_f,m_h] 
   -\frac{m_f^2}{v^2}s_\alpha^2 B_1[p^2;m_f,m_H] \notag\\[+5pt]
 & -\frac{m_f^2}{v^2} B_1[p^2;m_f,m_{G^0}]
   -\frac{m_{f'}^2+m_f^2}{v^2}B_1[p^2;m_f,m_{G^\pm}], \\[+10pt]
 (16\pi^2)\Pi_{ff,A}^{1\PI}[p^2]&=
   -\frac{m_{f}^2-m_{f'}^2}{v^2}B_1[p^2;m_f,m_{G^\pm}]
   +\frac{m_W^2}{v^2}(2B_1[p^2;m_{f'},m_W]+1)\notag\\[+5pt]
 & +2\frac{m_Z^2}{v^2}v_f a_f(2B_1[p^2;m_f,m_Z]+1),
 \end{align} 
where $v_f = I_f -2s_W^2 Q_f$, $a_f = I_f$ and $I_f^{}$ represents the third component of the isospin of a fermion $f$; i.e., $I_f^{}=+1/2$ $(-1/2)$ for $f=u$ $(d, e)$. 

 \subsection{Three point functions}
In this subsection, we use the simplified form for the three point function of the Passarino-Veltman as $C_i[X, Y, Z] \equiv C_i[p_1^2, p_2^2, q^2; m_X, m_Y, m_Z]$.
The 1PI diagram contributions for each form factor of the $hZZ$ and the $hWW$ couplings defined in Eq.~(\ref{eq:hvv_form}) are calculated as   
 \begin{align}  
 M_{hZZ,1}^{1\PI,F}[p_1^2,p_2^2,q^2] &= 
   \sum_{f}\frac{32m_Z^2 m_f^2 N_c^f}{16\pi^2v^3} c_\alpha \times \notag\\[+5pt]
 &\Big[ 
    \Big(\frac{1}{2}I_f^2-I_fQ_fs_W^2 + Q_f^2 s_W^4\Big)
  \big(2p_1^2C_{21} +2p_2^2C_{22} +4p_1\cdot p_2 C_{23} +2(D-2)C_{24} \notag\\[+5pt]
 & +(3p_1^2+p_1\cdot p_2)C_{11} +(3p_1\cdot p_2 +p_2^2)C_{12} 
   +(p_1^2 +p_1\cdot p_2)C_0\big) \notag\\[+5pt]
 & +(I_fQ_fs_W^2-Q_f^2s_W^4)
   \big( p_1^2 C_{21} +p_2^2C_{22} +2p_1\cdot p_2 C_{23} +DC_{24} \notag\\[+5pt]
 & +m_f^2 C_0 +(p_1^2+p_1\cdot p_2)C_{11} +(p_1\cdot p_2 +p_2^2)C_{12} \Big)\Big]
  [f,f,f], \\[+5pt]
 M_{hZZ,2}^{1\PI,F}[p_1^2,p_2^2,q^2] &=
   \sum_{f}\frac{4m_Z^2 m_f^2 N_c^f}{16\pi^2v^3} c_\alpha \times \notag\\[+5pt]
 & \Big[ (v_{f}^2 + a_{f}^2) \big(
    4C_{23} +C_{11} +3C_{12} +C_0 \big) 
  + (v_{f}^2 + a_{f}^2) \big(C_{12} - C_{11} \big) \Big][f,f,f],
 \end{align}
 \begin{align}
  M_{hZZ,3}^{1\PI,F}[p_1^2,p_2^2,q^2] &=
    -\sum_{f}\frac{8m_Z^2 m_f^2 N_c^f}{16\pi^2v^3} c_\alpha v_{f}a_{f}
   [C_{11} + C_{12} +C_0 ][f,f,f], \\[+5pt]
 M_{hWW,1}^{1\PI,F}[p_1^2,p_2^2,q^2]&= 
   \sum_{f}\frac{4m_W^2 m_f^2 N_c^f}{16\pi^2v^3} c_\alpha 
\Big[ 
   2p_1^2C_{21} +2p_2^2C_{22} +4p_1\cdot p_2 C_{23} +(2D-4)C_{24} \notag\\[+5pt]
 & +(3p_1^2+p_1\cdot p_2)C_{11} +(3p_1\cdot p_2 +p_2^2)C_{12} 
   +(p_1^2 +p_1\cdot p_2)C_0\big) \big]
  [f,f',f], \\[+5pt]
 M_{hWW,2}^{1\PI,F}[p_1^2,p_2^2,q^2]&=
   -\sum_{f}\frac{4m_W^2 m_f^2 N_c^f}{16\pi^2v^3} c_\alpha
  \big[
    4C_{23} +C_{11} +3C_{12} +C_0 \big][f,f',f], \\[+5pt]
 M_{hWW,3}^{1\PI,F}[p_1^2,p_2^2,q^2] &=
    -\sum_{f}\frac{4m_W^2 m_f^2 N_c^f}{16\pi^2v^3} c_\alpha
   [C_{11} + C_{12} +C_0 ][f,f',f],
 \end{align}

 \begin{align}
& (16\pi^2)M_{hZZ,1}^{1\PI,B}[p_1^2,p_2^2,q^2] =
 -2g^3 m_W c_W^2c_\alpha (D-1)B_0[q^2;m_W,m_W] \notag\\[+5pt]
 & -g g_Z^2 m_W s_W^4 c_\alpha (B_0[p_1^2;m_W,m_{G^\pm}] 
  + B_0[p_2^2;m_W,m_{G^\pm}) \notag\\[+5pt]
 &-\frac{g_Z^3 m_Z}{2}c_\alpha\left\{
c_\alpha^2 (B_0[p_1^2;m_Z,m_h] + B_0[p_2^2;m_Z,m_h]) 
   + s_\alpha^2(B_0[p_1^2;m_Z^{},m_H^{}] + B_0[p_2^2;m_Z^{},m_H^{}])\right\}
  \notag\\[+5pt]
 &+\frac{g_Z^2}{2}(c_{2W}^{})^2 \lambda_{hG^+G^-}B_0[q^2;m_{G^+},m_{G^+}]
  +\frac{3g_Z^2}{2}c_\alpha^2 \lambda_{hhh} B_0[q^2;m_h^{},m_h^{}]
 \notag\\[+5pt]
 &+g_Z^2 c_\alpha s_\alpha \lambda_{hhH} B_0[q^2;m_h^{},m_H^{}] 
  +\frac{g_Z^2}{2} s_\alpha^2 \lambda_{hHH} B_0[q^2;m_H^{},m_H^{}] 
  +\frac{g_Z^2}{2}  \lambda_{hG^0 G^0} B_0[q^2;m_{G^0},m_{G^0}] \notag\\[+5pt]
 &+2g^3 m_W c_W^2 c_\alpha C_{VVV}^{hVV,1}[W,W,W]
  -2g_Z^2 g m_W^3 s_W^4 c_\alpha C_0[W,G^\pm,W] \notag\\[+5pt]
 & +g^3 m_W s_W^2 c_\alpha 
   \Big\{(C_{VVS}^{hVV,1}[W,W,G^\pm] + C_{SVV}^{hVV,1}[G^\pm,W,W] ) \notag\\[+5pt]
 & - \frac{c_{2W}}{c_W^2} (C_{24}[W,G^\pm,G^\pm] +C_{24}[G^\pm,G^\pm,W])\Big\}
  - g_Z^3 m_Z^3 c_\alpha\{c_\alpha^2 C_0[Z,h,Z]+s_\alpha^2 C_0[Z,H,Z]\}
  \notag\\[+5pt]
 & + \frac{g_Z^3}{2} m_Z c_\alpha \big\{ 
   c_\alpha^2(C_{24}[Z,h,G^0] +C_{24}[G^0,h,Z]) 
  + s_\alpha^2 (C_{24}[Z,H,G^0] +C_{24}[G^0,H,Z]) \big\}
 \notag\\[+5pt]
 &+2g_Z^2 m_Z^2\{ 3c_\alpha^2 \lambda_{hhh}C_0[h,Z,h]
                             +s_\alpha^2 \lambda_{hHH} C_0[H,Z,H] \} \notag\\[+5pt]
 &+2\lambda_{hhH}g_Z^2 m_Z^2c_\alpha s_\alpha \{ C_0[h,Z,H] 
                             + C_0[H,Z,h] \} 
 -2g^3 m_W c_W^2 c_\alpha C_{24}[c^\pm,c^\pm,c^\pm]
 \notag\\
 & +2g_Z^2 m_W^2 s_W^4 \lambda_{hG^+G^-}( m_W^2 s_W^4 C_0[G^\pm,W,G^\pm] 
  - (c_{2W})^2 C_{24}[G^\pm,G^\pm,G^\pm] ) \notag\\[+5pt]
 &-2g_Z^2 \big\{ 
 3\lambda_{hhh} c_\alpha^2 C_{24}[h,G^0,h]
  +\lambda_{hhH} c_\alpha s_\alpha (C_{24}[h,G^0,H] +C_{24}[H,G^0,h])
  \notag\\[+5pt]
 &  +\lambda_{hHH}s_\alpha^2 C_{24}[H,G^0,H]  
    +\lambda_{hG^0G^0} (c_\alpha^2 C_{24}[G^0,h,G^0]
                           +s_\alpha^2 C_{24}[G^0,H,G^0])\Big\}, 
 \end{align}

 \begin{align}
& (16\pi^2)M_{hWW,1}^{1\PI,B}[p_1^2,p_2^2,q^2]=
   -g^3 m_W (D-1)\{c_\alpha B_0(p^2;W,W)
  + s_\alpha B_0(p^2;Z,Z)\} \notag\\[+5pt]
 &-\frac{g^3}{2}m_Wc_\alpha^{} \{c_\alpha^2 (B_0[p_1^2;W,h] +B_0[p_2^2; W,h]) 
  + s_\alpha^2 (B_0[p_1^2;W,H] +B_0[p_2^2; W,H])\} 
   \notag\\[+5pt]
 &-\frac{g^3}{2}s_W^2m_W c_\alpha \left\{
   \frac{s_W^2}{c_W^2}(B_0[p_1^2;Z,G^\pm] +B_0[p_2^2; Z,G^\pm])
  + (B_0[p_1^2;\gamma,G^\pm] +B_0[p_2^2; \gamma,G^\pm])\right\} \notag\\[+5pt]
 &+ \frac{g^2}{2}\big\{
   \lambda_{hG^+G^-} B_0[p^2,G^\pm,G^\pm]
  +3\lambda_{hhh} c_\alpha^2 B_0[p^2;h,h]
  +\lambda_{hHH} s_\alpha^2 B_0[p^2;H,H] \notag\\[+5pt]
 &+2\lambda_{hhH} s_\alpha c_\alpha B_0[p^2;h,H]
  +\lambda_{hG^0G^0} B_0[p^2;G^0,G^0]\big\} \notag\\[+5pt]
 &+ g^3 m_W c_W^2 c_\alpha C_{VVV}^{hVV,1}[W,Z,W]
  + e^2 g m_W c_\alpha C_{VVV}^{hVV,1}[W,\gamma,W]
  + g^3 m_W c_\alpha C_{VVV}^{hVV,1}[Z,W,Z] \notag\\[+5pt]
 &-\frac{g^3}{2}m_W s_W^2 c_\alpha C_{VVS}^{hVV,1}[W,Z,G^\pm]
  +\frac{e g^2}{2}m_W s_W c_\alpha C_{VVS}^{hVV,1}[W,\gamma, G^\pm] \notag\\[+5pt] 
 &-g^3 m_W^3 c_\alpha\{c_\alpha^2C_0[W,h,W] +s_\alpha^2C_0[W,H,W] \}
  -g^2 g_Z m_Z^3 s_W^4 c_\alpha C_0[Z,G^+,Z] \notag\\[+5pt]
 &-\frac{g^3}{2}m_W s_W^2 c_\alpha (C_{SVV}^{hVV,1}[G^\pm,Z,W] 
  - C_{SVV}^{hVV,1} [W,\gamma,G^\pm]) \notag\\[+5pt]
 &+\frac{g^3}{2}m_W c_\alpha \left\{ (c_\alpha^2C_{24}[W,h,G^\pm]
                              + s_\alpha^2C_{24}[W,H,G^\pm]) \} 
  +\frac{s_W^2}{c_W^2} (C_{24}[Z,G^+,G^0] +C_{24}[G^0,G^\pm,Z])\right\}
 \notag\\[+5pt]
 &+g^2 m_W^2\left\{
   \lambda_{hG^+G^-} \frac{s_W^4}{c_W^2} C_0[G^\pm,Z,G^\pm] 
  +\lambda_{hG^+G^-} s_W^2 C_0[G^\pm,\gamma,G^\pm] 
  +6\lambda_{hhh}  c_\alpha^2 C_0[h,W,h] \right\}\notag\\[+5pt]
 &+2\lambda_{hhH} g^2 m_W^2 c_\alpha s_\alpha (C_0[h,W,H] +C_0[H,W,h])
  +2\lambda_{hHH} g^2 m_W^2 (s_\alpha)^2 C_0[H,W,H] \notag\\[+5pt]
 &+\frac{g^3}{2}m_W c_h \{ s_\alpha^2C_{24}[G^\pm,h,W]  
                               +c_\alpha^2C_{24}[G^\pm,H,W] \}\notag\\[+5pt]
 &-g^3m_W c_W^2 c_\alpha C_{24}(c^\pm,c^0,c^\pm)  
  -e^2 gm_W  s_\alpha C_{24}(c^\pm,c^\gamma,c^\pm)  
  -g^3m_W c_\alpha C_{24}(c^0,c^\pm,c^0)
  \notag\\[+5pt]
 &-\lambda_{hG^+G^-} g^2 \{ c_\alpha^2C_{24}[G^\pm,h,G^\pm]  
                         +s_\alpha^2C_{24}[G^\pm,H,G^\pm] +C_{24}[G^\pm,G^0,G^\pm] \}
 \notag\\[+5pt]
 &-6\lambda_{hhh}g^2 c_\alpha^2 C_{24}[h,G^\pm,h]
  -2\lambda_{hhH}g^2 c_\alpha s_\alpha ( C_{24}[h,G^\pm,H] + C_{24}[H,G^\pm,h] )
 \notag\\[+5pt]
 &-2\lambda_{hHH}g^2 (s_\alpha)^2 C_{24}[H,G^\pm,H]   
  -2\lambda_{hG^0G^0}g^2  C_{24}[G^0,G^\pm,G^0],   
 \end{align}

\begin{align}
 & (16\pi^2)M_{hZZ,2}^{1\PI,B}[p_1^2,p_2^2,q^2] =
 g^3 m_W^{}c_\alpha^{}\big\{
     2 c_W^2 C_{VVV}^{hVV,2}[W,W,W] 
   + s_W^2 C_{VVS}^{hVV,2}[W,W,G^\pm] \notag\\[+5pt]
 & + s_W^2 (-C_{23} +2C_0)\big\} [G^\pm,W,W] 
   -g^3 m_W \frac{s_W^2 c_{2W}}{c_W^2} c_\alpha \left\{
   C_{VSS}^{hVV,2} [W,G^\pm,G^\pm]
   +C_{23}[G^\pm,G^\pm,W] \right\} \notag\\[+5pt]
 & - g_Z^3 m_Z^{}c_\alpha^{}\Big\{
  s_W^2c_{2W}^{} (C_{VSS}^{hVV,2}[W,G^\pm,G^\pm]  + C_{23}[G^\pm,G^\pm,W] )
 \notag\\[+5pt]
 & - c_\alpha^2(C_{VSS}^{hVV,2}[Z,h,G^0]  + C_{23}[G^0,h,Z] ) \Big\}
   +\frac{g_Z^3}{2} m_Z c_\alpha s_\alpha^2 \left\{
  C_{VSS}^{hVV,2} [Z,H,G^0]+C_{23}[G^0,H,Z] \right\} \notag\\[+5pt]
 & - 2g^3 m_W c_W^2 c_\alpha
   C_{12}[c^{\pm},c^{\pm},c^{\pm}]
   - 2 g_Z^2 \Big\{c_{2W}^2 \lambda_{hG^+G^-}  C_{1223}[G^\pm,G^\pm,G^\pm]
 \notag\\[+5pt]
 & + 3 c_\alpha^2 \lambda_{hhh}  C_{1223}[h,G^0,h] 
   + c_\alpha s_\alpha \lambda_{hhH}  \left\{
     C_{1223}[h,G^0,H] +C_{1223}[H,G^0,h] \right\} \notag\\[+5pt]  
 & + s_\alpha^2 \lambda_{hHH}  C_{1223}[H,G^0,H]
   + \lambda_{hG^0G^0} \left\{ c_\alpha^2   C_{1223}[G^0,h,G^0] 
                +s_\alpha^2   C_{1223}[G^0,H,G^0] \right\}\Big\},
 \end{align}

 \begin{align}
 & (16\pi^2)M_{hWW,2}^{1\PI,B}[p_1^2,p_2^2,q^2] =
    g^3 m_W c_\alpha \Big\{
     c_W^2 C_{VVV}^{hVV,2}[W,Z,W]
    + s_W^2 C_{VVV}^{hVV,2}[W,\gamma,W] \notag\\[+5pt]
 &  + C_{VVV}^{hVV,2}[Z,W,Z] \Big\} 
 -  \frac{g^3}{2} m_W s_W^2 c_\alpha \left\{
      C_{VVS}^{hVV,2}[W,Z,G^\pm] + C_{VVS}^{hVV,2} [W,\gamma,G^\pm] 
    \right\}\notag\\[+5pt]
 &-  \frac{g^3}{2}  m_W s_W^2 c_\alpha \left\{
      (-C_{23} +2C_0) [G^\pm, Z,W] 
      (-C_{23} +2C_0) [G^\pm, \gamma,W] \right\}  \notag\\[+5pt]
 &+ \frac{g^3}{2}m_W^{} c_\alpha^{} \Big\{
    c_\alpha^3(C_{VSS}^{hVV,2}[W,h,G^\pm] + C_{23}[G^\pm,h,W])
  + s_\alpha^2(C_{VSS}^{hVV,2}[W,H,G^\pm] + C_{23}[G^\pm,H,W])\notag\\[+5pt]
 &+ \frac{s_W^2}{c_W^2}(C_{VSS}^{hVV,2}[Z,G^\pm,G^0] + C_{23}[G^0,G^\pm,Z])
  \Big\}  \notag\\[+5pt]
 & - g^3 m_W^{} c_\alpha\Big\{
   c_W^2 C_{12}[c^{\pm},c^0,c^{\pm}]
   +s_W^2C_{12}^{}
   [c^{\pm},c^\gamma,c^{\pm}]
  + C_{12}^{}
   [c^0,c^{\pm},c^0] \Big\}\notag\\[+5pt]
 &-  g^2 \lambda_{hG^+G^-} \left\{c_\alpha^2 C_{1223}[G^\pm,h,G^\pm]
                           +s_\alpha^2 C_{1223}[G^\pm,H,G^\pm]
                           + C_{1223}[G^\pm,G^0,G^\pm] \right\} \notag\\[+5pt]
 &- 6g^2 \lambda_{hhh} c_\alpha^2 C_{1223}[h,G^\pm,h]
  - 2g^2 \lambda_{hHH} s_\alpha^2 C_{1223}[H,G^\pm,H] \notag\\[+5pt]
 &- 2g^2 \lambda_{hhH} c_\alpha s_\alpha\left\{
    C_{1223}[h,G^\pm,H] +C_{1223}[H, G^\pm,h] \right\}
  - 2g^2 \lambda_{hG^0G^0} C_{1223}[G^0,G^\pm,G^0],
 \end{align}
 \begin{align}
 M_{hZZ,3}^{1\PI,B}[p_1^2,p_2^1,q^2] = M_{hWW,3}^{1\PI,B}[p_1^2,p_2^2,q^2] = 0,
 \end{align}
 where 
 \begin{align}
 C_{VVV}^{hVV,1}[X,Y,Z] &= \big\{
    (6D-6)C_{24} + p_1^2( 2C_{21} + 3C_{11} +C_0)
  + p_2^2 (2C_{22} + C_{12})  \notag\\
& + p_1\cdot p_2 (4C_{23}+3C_{12} +C_{11} -4C_0) 
   \big\} [X,Y,Z], \\
 C_{VVS}^{hVV,1}[X, Y,Z] & = \big\{
   (D-1)C_{24} + p_1^2 (C_{21}+4C_{11} + 4C_0) + p_2^2 (C_{22} + 2C_{12}) \notag\\
 & + p_1\cdot p_2 (2C_{23} + 4C_{12} + 2C_{11} + 4C_{0}) \big\}
   [X,Y,Z], \\
 C_{SVV}^{hVV,1}[X,Y,Z] & = \big\{
   (D-1) C_{24} + p_1^2 (C_{21} - C_0) + p_2^2( C_{22} -2C_{12} +C_0) \notag \\
 & + 2p_1\cdot p_2 (C_{23} - C_{11}) \big\}[X,Y,Z], \\
 C_{VVV}^{hVV,2}[X,Y,Z]&=
 [C_{11}^{} + 9C_{12}^{} + 10C_{23}^{} + 5C_0^{}][X,Y,Z] \\[+5pt]
 C_{VVS}^{hVV,2}[X,Y,Z]&=
 [2C_{11}^{} -5C_{12}^{} -C_{23}^{} -2C_0^{}][X,Y,Z]\\[+5pt]
 C_{VSS}^{hVV,2}[X,Y,Z]&=
 [C_{1223}^{} + 2C_{11}^{} +2C_{0}^{}][X,Y,Z], 
 \end{align}
 and $C_{1223}= C_{12}+C_{23}$.

We give 1PI diagram contributions to $hf\bar{f}$ couplings, which are composed of following seven form factors,
 \begin{align}
 F_{hff}^{1\PI}[p_1^2,p_2^2,q^2]&= \Big\{
  F_{hff,S}^{1\PI} + \gamma_5 F_{hff,P}^{1\PI} +\slashed{p}_1F_{hff,V1}^{1\PI}
 + \slashed{p}_2F_{hff,V2}^{1\PI} + \slashed{p}_1 \gamma_5 F_{hff,A1}^{1\PI}
 + \slashed{p}_2 \gamma_5 F_{hff,A2}^{1\PI} \notag\\[+5pt]
&+ \slashed{p}_1 \slashed{p}_2 F_{hff,T}^{1\PI}
 + \slashed{p}_1\slashed{p}_2 \gamma_5 F_{hff,TP}^{1\PI} \Big\}[p_1^2,p_2^2,q^2].
 \end{align}
Each part is calculated as
 \begin{align}
 & (16\pi^2)F_{hff,S}^{1\PI}[p_1^2,p_2^2,q^2]=
  -4c_\alpha \frac{m_f^{}}{v} \left\{
    \frac{m_Z^2}{v^2}(v_f^2 - a_f^2) C_{FVF}^{hff,S}[f,Z,f]
  + (Q_f e)^2 C_{FVF}^{hff,S}[f,Z,f] \right\} \notag\\[+5pt]
 & +\frac{m_f^3}{v^3}c_\alpha^{}\left\{
     c_\alpha^2C_{FSF}^{hff,S}[f,h,f]
   + s_\alpha^2 C_{FSF}^{hff,S}[f,H,f] 
   - c_\alpha C_{FSF}^{hff,S}[f,G^0,f] \right\}\notag\\[+5pt]
 &  -2c_\alpha \frac{m_f m_{f'}^2}{v^3}C_{FSF}^{hff,S}
     [f',G^\pm,f'] 
    -8c_\alpha \frac{m_Z^4}{v^3}m_f (v_f^2 - a_f^2)C_0[Z,f,Z] \notag\\[+5pt]
 & -2\frac{m_f^2}{v^2}\left\{
     3c_\alpha^2 \lambda_{hhh} C_0[h,f,h]
   + s_\alpha^2 \lambda_{hHH} C_0[H,f,H] 
   + c_\alpha s_\alpha \lambda_{hhH}(C_0[h,f,H]+C_0[H,f,h]) \right\}
 \notag\\[+5pt]
 &  + 2\frac{m_f^2}{v^2}\left\{
      \lambda_{hG^0G^0} 
   \frac{m_f^{}}{v} C_0[G^0,f,G^0] 
    + \lambda_{hG^+G^-} 
   \frac{m_{f'}^{}}{v} C_0[G^\pm,f',G^\pm] \right\}  \notag\\[+5pt]
 &  -c_\alpha \frac{m_W^2 m_f}{v^3}(C_{SFV}^{hff,S}[G^\pm,f',W] +C_{VFS}^{hff,S}[W,f',G^\pm] )
\notag\\[+5pt]
 &  -c_\alpha \frac{m_Z^2 m_f}{2v^3}(C_{SFV}^{hff,S}[G^0,f,Z] +C_{VFS}^{hff,S}[Z,f,G^0]),
 \end{align}
 \begin{align}
  F_{hff,P}^{1\PI}[p_1^2,p_2^2,q^2] &=
   -\frac{1}{16\pi^2}\frac{m_f^{}}{v^2}c_\alpha^{}\Big\{
   m_W^2 (  C_{VFS}^{hff,T}[W, f', G^\pm] 
         -  C_{SFV}^{hff,T}[G^\pm, f', W] ) \notag\\[+5pt]
  & +2I_f^{} v_f^{} m_Z^2 ( 
     C_{VFS}^{hff,T} [Z, f, G^0] - C_{SFV}^{hff,T} [G^0, f, Z]) \Big\},
 \end{align}
 \begin{align}
 & (16\pi^2)F_{hff}^{V1}[p_1^2,p_2^2,q^2]=
    \frac{m_f^4}{v^3} c_\alpha^{} \big\{
     c_\alpha^2 (C_0+2C_{11})[f, h, f]   
   + s_\alpha^2 (C_0+2C_{11})[f, H, f] \notag\\[+5pt]
 &  +  (C_0+2C_{11})[f, G^0, f] \big\} 
   +\frac{m_{f'}^2}{v} c_\alpha^{}
   \left( \frac{m_f^2}{v^2}+ \frac{m_{f'}^2}{v^2}\right)
      \left\{C_0+2C_{11}\right\}[f', G^\pm, f'] \notag\\[+5pt]
 &  +2\frac{c_\alpha^{}}{v^3}\left\{
     m_W^2m_{f'}^2 (C_0+2C_{11})[f', W, f'] 
    +m_Z^2m_f^2 (v_{f}^2+a_{f}^2)(C_0+2C_{11})[f, Z, f] \right\}\notag\\[+5pt]
 & +2 (Q_f e)^2 \frac{m_f^2}{v} c_\alpha
      \left\{C_0+2C_{11}\right\}[f, \gamma, f] 
   + \lambda_{hG^+G^-}
    \left(\frac{m_f^2}{v^2}+\frac{m_{f'}^2}{v^2}\right)
      \left\{C_0+C_{11}\right\}[G^\pm, f', G^\pm] \notag\\
 & + 2\frac{m_f^2}{v^2}\Big\{
    3 \lambda_{hhh} c_\alpha^2 (C_0+C_{11})[h, f, h] 
   +  \lambda_{hHH} s_\alpha^2 (C_0+C_{11})[H, f, H] \notag\\[+5pt]
 & +  \lambda_{hhH} s_\alpha^{} c_\alpha^{}
      (C_0+C_{11})\left\{[h, f, H] + [H,f,h] \right\} 
   +2 \lambda_{hG^0G^0} (C_0+C_{11})[G^0, f, G^0]\Big\} \notag\\[+5pt] 
 & -4c_\alpha \frac{m_Z^4}{v^3}(v_{f}^2+a_{f}^2)
      \left\{C_0+C_{11}\right\}[Z, f, Z]
   -4c_\alpha \frac{m_W^4}{v^3}
      \left\{C_0+C_{11}\right\}[W, f', W] \notag\\
 & -c_\alpha \frac{m_W^2}{v^2} \frac{m_{f'}^2}{v}
      \left\{(2C_0+C_{11})[W, f', G^\pm] 
      -(C_0-C_{11})[G^\pm, f', W] \right\}\notag\\
 &  -2I_f c_\alpha \frac{m_Z^2}{v^2} \frac{m_f^2}{v} a_{f}
      \left\{(2C_0+C_{11})[Z, f, G^0] 
     -  (C_0-C_{11})[G^0, f, Z]\right\} ,
 \end{align}
 \begin{align}
 &(16\pi^2) F_{hff,V2}^{1\PI}[p_1^2,p_2^2,q^2] =
    \frac{m_f^4}{v^3} c_\alpha^{} \Big\{
   c_\alpha^2(C_0+2C_{12})[f, h, f]  +s_\alpha^2 (C_0+2C_{12})[f, H, f] 
  \notag\\[+5pt]
 & +( C_0+2C_{12}[f, G^0, f]\Big\}
   +\frac{m_{f'}^2}{v} c_\alpha
   \left( \frac{m_f^2}{v^2}+ \frac{m_{f'}^2}{v^2}\right)
      \left\{C_0+2C_{12}\right\}[f', G^\pm, f'] \notag\\[+5pt]
 & + 2\frac{c_\alpha^{}}{v^3}\left\{
    m_W^2 m_{f'}^2 (C_0+2C_{12})[f', W, f'] 
   + m_Z^2m_f^2(v_{f}^2 + a_{f}^2) (C_0+2C_{12})[f, Z, f] \right\}\notag\\[+5pt]
 & +2 (Q_f e)^2 \frac{m_f^2}{v} c_\alpha
      \left\{C_0+2C_{12}\right\}[f, \gamma, f]
   + \lambda_{hG^+G^-}
    \left(\frac{m_f^2}{v^2}+\frac{m_{f'}^2}{v^2}\right)
      C_{12}[G^\pm, f', G^\pm] \notag\\[+5pt]
 & + 2\frac{m_f^2}{v^2}\Big\{ 
     3 \lambda_{hhh} c_\alpha^2  C_{12}[h, f, h] 
   +   \lambda_{hHH} s_\alpha^2  C_{12}[H, f, H] 
   +  \lambda_{hG^0G^0} C_{12}[G^0, f, G^0] \notag\\[+5pt]
 & +  \lambda_{hhH} s_\alpha c_\alpha (C_{12}[h, f, H] + [H,f,h])  \Big\}
   -4 \frac{c_\alpha^{}}{v^3} \left\{
      m_Z^4(v_{f}^2+a_{f}^2) C_{12}[Z, f, Z]
    + m_W^4 C_{12}[W, f', W] \right\} \notag\\[+5pt]
 &  - c_\alpha \frac{m_W^2}{v^2} \frac{m_{f'}^2}{v} \left\{
      (2C_0+C_{12})[W, f', G^\pm] - (C_0-C_{12})[G^\pm, f', W] \right\}
   \notag\\[+5pt]
 & -2I_f c_\alpha \frac{m_Z^2}{v^2} \frac{m_f^2}{v} a_{f}\left\{
       (2C_0+C_{12})[Z, f, G^0] - (C_0-C_{12})[G^0, f, Z] \right\},
 \end{align}
 \begin{align}
 &  F_{hff,A1}^{1\PI}[p_1^2,p_2^2,q^2] =\frac{1}{16\pi^2}\Bigg\{
    2c_\alpha^{}\frac{m_W^2}{v^3}\left\{
     2m_W^2 (C_0^{} + C_{11}^{})[W,f',W]
    - m_{f'}^2 (C_0^{} + 2C_{11}^{})[f',W, f'] \right\} \notag\\[+5pt]
 & + 4c_\alpha^{} \frac{m_Z^2}{v^3}v_f^{} a_f^{} \left\{
     2m_Z^2 (c_0^{} + C_{11}^{})[Z,f,Z]
   - m_f^2 (C_0^{} + 2C_{11}^{})[f,Z,f] \right\} \notag\\[+5pt]
 & + c_\alpha \frac{m_W^2}{v^2} \frac{m_{f'}^2}{v}
      \left\{2C_0+C_{11}\right\}[W, f', G^+] 
   - c_h \frac{m_W^2}{v^2} \frac{m_{f'}^2}{v}
      \left\{C_0-C_{11}\right\}[G^+, f', W] \notag\\
 & + c_\alpha^{} \frac{m_W^2}{v^2} \frac{m_{f'}^2}{v}\left\{
    (2C_0^{} + C_{11}^{})[W,f',G^\pm] -(C_0^{}-C_{11}^{})[G^\pm, f',W]\right\}
 \notag\\[+5pt]
 & + c_\alpha^{} I_f^{}v_f^{} \frac{m_Z^2}{v^2} \frac{m_f^2}{v}\left\{
    (2C_0^{} + C_{11}^{})[Z,f,G^0] -(C_0^{}-C_{11}^{})[G^0, f,Z]\right\}
 \notag\\[+5pt]
 & + \left( \frac{m_f^2}{v^2} - \frac{m_{f'}^2}{v^2} \right)
  \left\{ \frac{m_{f'}^2}{v} c_\alpha^{} (C_0^{} + 2C_{11}^{})[f', G^\pm, f']
  + \lambda_{hG^+G^-} (C_0^{} + C_{11}^{})[G^\pm, f', G^\pm] \right\},
 \end{align}

 \begin{align}
 & F_{hff,A2}^{1\PI}[p_1^2,p_2^2,q^2]= \frac{1}{16\pi^2} \Bigg\{
  2c_\alpha^{}\frac{m_W^2}{v^3}\left\{
    2m_W^2C_{12}^{}[W,f',W] - m_{f'}^2 (C_0^{}+2C_{12}^{})[f',W,f'] \right\}
  \notag\\[+5pt]
 & + 4c_\alpha^{}\frac{m_Z^2}{v^3}v_f^{} a_f^{}\left\{
   2m_Z^2 C_{12}[Z,f,Z] - m_f^2 (C_0^{}+2C_{12}^{})[f,Z,f] \right\}
   \notag\\[+5pt]
 & +c_\alpha^{}\frac{m_W^2}{v^2}\frac{m_{f'}^2}{v}\left\{
   (2C_0^{}+C_{12}^{})[W,f',G^\pm] - (C_0^{} - C_{12}^{})[G^\pm,f',W] \right\}
   \notag\\[+5pt]
 & +2 I_f^{}v_f^2 c_\alpha^{} \frac{m_Z^2}{v^2}\frac{m_f^2}{v}\left\{
  (2C_0^{} + C_{12}^{})[Z,f,G^0] + (C_0^{} - C_{12}^{})[G^0,f,Z] \right\}
    \notag\\[+5pt]
 & + \left(\frac{m_f^2}{v^2} - \frac{m_{f'}^2}{v^2}\right)
    \left(\frac{m_{f'}^2}{v}c_\alpha^{}(C_0^{}+2C_{12}^{})[f',G^\pm,f']
       + \lambda_{hG^+G^-}^{} C_{12}^{}[G^\pm, f',G^\pm] \right)
  \Bigg\},
 \end{align}
 \begin{align}
 & F_{hff,T}^{1\PI}[p_1^2,p_2^2,q^2] =\frac{1}{16\pi^2}\Bigg\{
  \frac{m_f^3}{v^3}c_\alpha^{}\left\{
    c_\alpha^2 (C_{11}^{}-C_{12}^{})[f,h,f]
  + s_\alpha^2 (C_{11}^{} - C_{12}^{})[f,H,f] \right\} \notag\\[+5pt]
 & \frac{m_f^2}{v^3}c_\alpha^{}\left\{
  m_f (C_{11}^{} - C_{12}^{})[f,G^0,f] 
  - 2m_{f'}^{} (C_{11}^{} - C_{12}^{}) [f', G^\pm, f'] \right\} \notag\\[+5pt]
 & + c_\alpha \frac{m_W^2}{v^2} \frac{m_f}{v} \left\{
     (2C_0 +2C_{11} -C_{12})[W, f', G^+] 
     (C_0 +C_{11} -2C_{12} )[G^+, f', W] \right\} \notag\\[+5pt]
 & +c_\alpha \frac{m_Z^2}{2v^2}\frac{m_f}{v} \left\{
     (2C_0 +2C_{11} -C_{12})[Z, f, G^0]       %
   + (C_0 +C_{11} -2C_{12} )[G^0, f, Z] \right\}
  \Bigg\},     
 \end{align}

 \begin{align}
 & F_{hff,TP}^{1\PI}[p_1^2,p_2^2,q^2] \notag\\[+5pt]
 & =\frac{1}{16\pi^2}\Bigg\{
     c_\alpha \frac{m_W^2}{v^2} \frac{m_f}{v} \Big\{
     (2C_0 +2C_{11} -C_{12})[W, f', G^+] 
   -  (C_0 +C_{11} -2C_{12})[G^+, f', W] \Big\}\notag\\[+5pt]
 & +2I_f^{}v_f^{} c_\alpha \frac{m_Z^2}{v^2}\frac{m_f}{v}\Big\{
      (2C_0 +2C_{11} -C_{12})[Z, f, G^0] 
    - (C_0 +C_{11} -2C_{12} )[G^0, f, Z] \Big\} \Bigg\},
 \end{align}
where
 \begin{align}
 C_{FVF}^{hff,S}[X,Y,X]&= \{p_1^2(C_{11}+C_{21})+p_2^2(C_{12}+C_{22}) 
        +p_1\cdot p_2(C_{11}+C_{12}+2C_{23}) \notag\\[+5pt]
             &+4 C_{24} -1 +m_X^2 C_0 \}[X,Y,X], \\[+5pt]
 C_{FSF}^{hff,S}[X,Y,X]&= \{ p_1^2(C_{11}+C_{21}) +p_2^2(C_{12}+C_{22}) 
  +2p_1\cdot p_2(C_{12}+C_{23}) \notag\\[+5pt]
             &+4C_{24}-\frac{1}{2} +m_X^2 C_0\}[X,Y,X], \\[+5pt]
 C_{SFV}^{hff,S}[X,Y,Z]&= \{p_1^2(C_{21}-C_0) +p_2^2(C_{22}-C_{12}) 
  +2p_1\cdot p_2(C_{23}-C_{12}) +4C_{24} - \frac{1}{2} \}[X,Y,Z] \\[+5pt]
 C_{VFS}^{hff,S}[X,Y,Z]&= \{p_1^2(C_{21}+3C_{11}+2C_0) +p_2^2(C_{22}+2C_{12}) \notag\\[+5pt]
 & +2p_1\cdot p_2(C_{23}+C_{12}+2C_{11}+2C_0) +4C_{24} - \frac{1}{2} \}[X,Y,Z],\\
  C_{VFS}^{hff,T}[X,Y,Z]&= \big\{ p_1^2(3C_{11}+2C_0+C_{21}) +p_2^2(2C_{12}+C_{22}) 
          \notag\\[+5pt]
               & +2p_1\cdot p_2(C_{12}+C_{23}+2C_{11}+2C_0) 
                 + DC_{24} \big\}[X,Y,Z], \\[+5pt]
  C_{SFV}^{hff,T}[X,Y,Z]&=\left\{ p_1^2(C_{21}-C_0) +p_2^2(C_{22}-C_{12})
           +2p_1\cdot p_2(C_{23}-C_{12}) +DC_{24}  \right\}[X,Y,Z].
 \end{align}

 \section{Decay rates of one-loop induce processes}\label{sec:decay rate}
We also list decay rate formlue of 1 loop induce processes; i.e., $h\rightarrow \gamma\gamma$~\cite{extended_hgammagamma_hZg} , $h \rightarrow \gamma Z$~\cite{extended_hgammagamma_hZg}  and $h \rightarrow gg$~\cite{hgg}.
\begin{align}
 \Gamma(h\rightarrow\gamma\gamma)&=
   \frac{\sqrt{2}c_\alpha^2G_F \alpha_{em}^2 m_h^3}{256\pi^3}
    \left|  \sum_f  Q_f^2 N_c^f I_f(m_h) +I_W(m_h)
    \right|^2 , \\
 \Gamma(h\rightarrow \gamma Z)&=
   \frac{\sqrt{2}c_\alpha^2G_F \alpha_{em}^2 m_h^3}{128 \pi^3}
   \left(1-\frac{m_Z^2}{m_h^2}\right)^3
   \left|\sum_f  Q_F N_f^c v_f J_f(m_h) + J_W(m_h) \right|^2, \\
 \Gamma(h\rightarrow gg)&=
   \frac{\sqrt{2}c_\alpha^2G_F \alpha_s^2 m_h^3}{128 \pi^3}
    \left|\sum_f I_f(m_h) \right|^2, 
  \end{align}
 where
  \begin{align}
  I_f(m_h)& = \frac{8m_f^2}{m_h^2}\left\{
   1+ \left(2m_f^2-\frac{m_h^2}{2}\right)C_0
  [0,0,m_h^2;m_f,m_f,m_f] \right\}, \\
 I_W(m_h)& = \frac{2m_W^2}{m_h^2} \left\{
  6+\frac{m_h^2}{m_W^2} +(12m_W^2 -6m_h^2)C_0
    [0,0,m_h^2;m_W,m_W,m_W] \right\}, \\
 J_f(m_h)& = - \frac{8m_f^2}{s_W^{}c_W^{}(m_h^2 - m_f^2)}
  \left(1+ \frac{1}{2}(4m_f^2 - m_h^2 +m_Z^2) 
   C_0^{}[0,m_Z^2,m_h^2; m_f^{},m_f^{},m_f^2] \right)\notag\\[+5pt]
 & + \frac{m_Z^2}{m_h^2 - m_Z^2} (B_0^{}[m_h^2;m_f^{},m_f^{}]
                           - B_0^{}[m_Z^2; m_f^{},m_f^{}] ) , \\
 J_W(m_h)& =
 \frac{2m_W^2}{s_W^{}c_W^{}(m_h^2 - m_Z^2)}\Bigg\{
2(s_W^2-3c_W^2) (m_h^2 - m_Z^2) C_0[0,m_Z^2,m_h^2;m_W^{},m_W^{},m_W^{}] 
 \notag\\[+5pt]
& +\left(  c_W^2\left(5 + \frac{m_h^2}{2m_W^2} \right)
        - s_W^2\left(1 + \frac{m_h^2}{2m_W^2} \right) \right) 
  \Big( 1 + 2m_W^2 C_0^{}[0,m_Z^2,m_h^2; m_W^{},m_W^{},m_W^{}] \notag\\[+5pt]
&     + \frac{m_Z^2}{m_h^2 - m_W^2}
   (B_0^{}[m_h^2; m_W^{},m_W^{}] - B_0[m_Z^2; m_W^{},m_W^{}] )\Big) 
  \Bigg\}. 
  \end{align}



\end{document}